%% file: main.tex
\documentclass[a4paper,UKenglish,cleveref, autoref, thm-restate]{lipics-v2021}
\nolinenumbers
\usepackage{amsmath}
\usepackage{amsfonts}
\usepackage{amssymb}
\usepackage{amsthm}
\usepackage{xspace}
\usepackage{stmaryrd}
\usepackage{amscd}
\usepackage{verbatim}
\usepackage{yfonts}
\usepackage{versions}

\usepackage{calc}

\usepackage{hyperref}
\usepackage{pdfpages} 
  \usepackage{ulem}

\usepackage{float} 

 \usepackage{color}
\usepackage{listings}
\usepackage{graphicx}
\usepackage{wrapfig}

\usepackage{pgf}
\usepackage{tikz}
\usetikzlibrary{arrows,automata,shapes}
\usepackage{mathtools} 

\newcommand{\Nat}{\mathbb{N}}

\newcommand{\ksurdeux}{\floor{\frac{k}{2}}}
\newcommand{\motnouveau}[1]{\textit{#1}}
\newcommand{\etal}{\textit{et al.}}
\newcommand{\Exe}{\sigma}

\newcommand{\quickset}[1]{\left\lbrace #1 \right\rbrace}

\newcommand{\bigconcchoose}[1]{\def\bigconcsize{}%
  \ifx#1\displaystyle
    \let\bigconcsize\Big
  \else
    \ifx#1\textstyle
      \let\bigconcsize\big
    \fi
  \fi#1}

\newif\ifcommentaires\commentairestrue

\newenvironment{articlelong}{}{}
\excludeversion{articlelong}

\newenvironment{articlecourt}{}{}
\includeversion{articlecourt}

\excludeversion{mainarticle}

\newcommand{\local}{{\sc Local}}
\newcommand{\lcl}{{\sc Lcl}}

\newcommand{\segment}[2]{\llbracket #1, #2 \rrbracket}

\newcommand{\floor}[1]{\lfloor #1 \rfloor}
\newcommand{\ceil}[1]{\lceil #1 \rceil}
\newcommand{\KAmoinsUn}{k-1}
\newcommand{\KA}{k}

\newcommand{\welld}[1]{well\_defined( #1 )}
\newcommand{\bc}[1]{branch\_coherence( #1 )}

\newcommand{\bcd}[2]{branch\_coherence\_down( #1 , #2 )}

\title{Making Self-Stabilizing Algorithms for any Locally Greedy Problem}

\author{Johanne Cohen}{Université Paris-Saclay, CNRS, LISN, 91405, Orsay, France.}{johanne.cohen@lisn.upsaclay.fr}{https://orcid.org/0000-0002-9548-5260}{}
\author{Laurence Pilard}{LI-PaRAD,  UVSQ, Universit\'e Paris-Saclay, France,}{laurence.pilard@uvsq.fr}{https://orcid.org/0000-0002-1104-8216}{}
\author{Mikaël Rabie}{IRIF-CNRS, Universit\'e Paris Cité, France}{mikael.rabie@irif.fr}{}{}
\author{Jonas Sénizergues}{Université Paris-Saclay, CNRS, LISN, 91405, Orsay, France.}{jonas.senizergues@lri.fr}{}{}

\authorrunning{J. Cohen, L. Pilard, M. Rabie and  J. Sénizergues } 

\Copyright{Johanne Cohen, Laurence Pilard, M. Rabie and  J. Sénizergues } 



\category{} 

\relatedversion{} 

\begin{document}

\maketitle

\input{introduction}



\input{algo}

\input{Explanation}

\input{proof}

\section{Conclusion}

This work provides a self-stabilizing algorithm under the Gouda daemon for any locally mendable problem by first introducing an explicit algorithm to compute a $(k,k-1)$-ruling set. This construction generalises well to probabilistic daemons if stationary rules and rule \textbf{Become Leader} have some probability smaller than $1$ to be activated. This algorithm permits building up distance-$k$ colorings, which helps solve greedy and mendable problems by simulating the LOCAL model. {In the case of constant bounded degree $\Delta$, our algorithm uses a constant memory.} We did not consider complexity questions. Considering a probabilistic daemon, an open question would be what complexities can be aimed, as our algorithm did not optimize this question at all. 

The presented algorithm for the ruling set should adapt well in the Byzantine case, as the influence of a Byzantine node is naturally confined by the algorithm. In such context, Distance-$K$ identifiers computed in Section~\ref{sec:distCol} would be unique at distance $K$ for nodes far enough from Byzantine nodes. It should be of interest to investigate this point. 



\bibliography{Bibliography}



\end{document}

%% file: introduction.tex
\begin{abstract}

We propose a way to transform synchronous distributed algorithms solving locally greedy and mendable problems into self-stabilizing algorithms in anonymous networks. Mendable problems are a generalization of greedy problems where any partial solution may be transformed -instead of completed- into a global solution:  every time we extend the partial solution we are allowed to change the previous partial solution up to a given distance. Locally here means that to extend a solution for a node, we need to look at a constant distance from it. 
 
In order to do this, we propose the first explicit self-stabilizing algorithm computing a $(k,k-1)$-ruling set (\textit{i.e.} a ``maximal independent set at distance $k$''). By combining multiple time this technique, we compute a distance-$K$ coloring of the graph. With this coloring we can finally simulate \local~model algorithms running in a constant number of rounds, using the colors as unique identifiers.

Our algorithms work under the Gouda daemon, which is similar to the probabilistic daemon: if an event should eventually happen, it will occur under this daemon.

\end{abstract}

\section{Introduction}

The greedy approach is often considered to solve a problem:
is it possible to build up a solution step by step by completing a partial solution? 
For example, in graph theory, one can consider the Maximal Independent Set (MIS) problem where you want to select a set of nodes such that no two chosen nodes are adjacent and any unselected node is a neighbor of a selected one.
To produce an MIS, a simple algorithm selects a node, rejects all its neighbors, and then repeats this operation until no node is left.
Another classical greedy algorithm is the one that produces a $(\Delta+1)$-coloring of a graph, where $\Delta$ is the maximum degree in the graph. Each time a node is considered, as it has at most $\Delta$ different colors in its neighborhood, we can always choose a different color for it to extend the current partial solution. Observe that most graphs admit a $\Delta$-coloring, which cannot be found with this heuristic and that those solutions. We can also notice that the size of an MIS can be arbitrarily smaller than the size of a \textit{maximum} independent set.  \motnouveau{Greedy problems} are problems that can be solved using a greedy algorithm. 

\motnouveau{Mendable} problems have recently been introduced in~\cite{balliu2021local} as a generalization of greedy problems. In these problems, a solution can be found sequentially by producing the output of each node one after another, as it is the case in greedy problems. However, here, for each chosen node, it is possible to change the output of its neighborhood, but only up to some distance.
The set of mendable problems is larger than the set of greedy ones. For instance, the $4$-coloring of the grid is a mendable problem, but it cannot be solved greedily, as its maximal degree $\Delta$ is equal to $4$.

A more generalized way to consider MIS are \motnouveau{ruling sets}. Given a graph $G=(V, E)$, a \motnouveau{$(a,b)$-ruling set} is a subset $S\subset V$ such that the distance between any two nodes in $S$ is at least $a$, and any node in $V$ is at distance at most $b$ from some node in $S$. We denote by $dist(u,v)$ the distance between the two nodes $u$ and $v$ in the graph. When $S$ is a subset of $V$, $dist(u,S)$ is the smallest distance from $u$ to an element in $S$. In what follows, the concept of \motnouveau{ball} will play an important role. Formally,  the \motnouveau{ball} of radius $i$ and center $s$, $\mathcal B (s,i)$, is the set of nodes that are at distance at most $i$ from $s$. Observe that a ball of radius $a-1$ centered in a node of the ruling set $S$ contains only one node in $S$. 

In particular, a $(2, 1)$-ruling set is an MIS of $G$. A $(k,k-1)$-ruling set $S$ is a maximal independent set at distance $k$: all the elements of $S$ are at distance at least $k$ from each other,  every other node is at distance at most $k-1$ from $S$, and thus cannot be added.
Note that it is an MIS of $G^{k-1}$ (the graph with the same vertices as $G$, and with edges between two vertices if there are at distance $k-1$ or less from each other in $G$), and this problem can be greedily solved.
A $(k,k-1)$ ruling set can also be seen as a maximal distance-$k$ independent set where a  distance-$k$ independent set is a subset of nodes at distance at least $k$ from each other.

 A \motnouveau{distance-$K$ coloring} 
 of a graph $G=(V,E)$ is a mapping $\mathcal C: V\to \mathbb{N}$ such that $\forall u\neq v\in V^2$, $dist(u,v)\le K\Rightarrow \mathcal C(u)\neq \mathcal C(v)$.
A way to produce a distance-$K$ coloring is to partition $V$ into sets of nodes at distance at least $k>K$ from each other, \textit{i.e.} distance-$k$ independent sets, each one representing a color. One can construct such a partition sequentially by constructing a partition into $X\ge\Delta^k$ 
distance-$k$ independent sets  $\{S^{(i)}\}_{i\le X}$, where $S^{(i)}$ is a 
distance-$k$ independent set of $G$ maximal under the constraint that every node of the independent must be in $V\setminus\bigcup_{j<i}S^{(j)}$. These distance-$k$ independent set can be computed in a very similar way as $(k,k-1)$-ruling sets. Distance-$K$ coloring is a way to simulate \local\  algorithms: the colors can be used as constant size identifiers, as long as the simulated algorithm does run in less than $K$ rounds on this range of identifiers.

\subsection{Our Contribution}

In this paper, we provide the self-stabilizing algorithm that computes a $(k,k-1)$-ruling set under the Gouda daemon (Section~\ref{Sec:RulingSet}).
The algorithm detects when a leader can be added or  two leaders are too close. To that end, each node computes its distance to the leaders. If a node and its neighbors are at distance at least $k-1$ from the leaders, that node can try to add itself to the ruling set. If two leaders are too close, thanks to a clock system, a node in the middle of the path will eventually detect the problem and initiate the removal of the leaders from the set.
Thanks to the Gouda daemon, we ensure that not too many nodes will try to add themselves simultaneously and that the clock system will eventually detect collisions. On the other hand, we prove that a stable configuration can always be reached, and the Gouda daemon ensures that it ultimately happens (Section~\ref{sec:proof}).

In Section~\ref{sec:distCol}, by combining this algorithm $\Delta^k$ times, we partition the graph into distance-$k$ independent sets, which correspond to a distance-$K$ coloring for any $K<k$. This coloring allows us to consider nodes of each set sequentially to compute a solution to some greedy problem. 
In Section~\ref{sec:mending}, we present a solution allowing us to solve any $T$-mendable problem, where $T$ is a constant corresponding to the radius up to which we are permitted to change the output of a node. 
To that end, we use the fact that a \local~algorithm runs in $r$ rounds for some constant $r$, when a distance-$2T+1$ coloring is given. To do that, we compute a distance-$2T+1$ and a distance-$2r+1$ coloring. That way, each node can  access their neighborhood at the right distance and compute the output the \local~algorithm would have given in that situation.

\subsection{Related Work}

The notion of \motnouveau{checking locally} and its relationship with the idea of \motnouveau{solving locally} has been introduced by Naor and Stockmeyer in~\cite{naor1995can}. This work, along with Cole and Vishkin's algorithm that efficiently compute a 3-coloring of a ring~\cite{cole1986deterministic}, leads to the notion of \motnouveau{Locally Checkable Labelling problems} (\lcl) and the \local~model. Locally checkable problems are problems such that when the output is locally correct for each node, the global output is guaranteed to be correct too.
Coloring and MIS belong to that field. Ruling Sets are also \lcl~problems: to check locally that the solution is right, the distance to the set must be given in the output. The \local~model (see~\cite{suomela2013survey} for a survey) is a synchronous model that requires unique identifiers but does not impose any restriction on communication bandwidth or computation complexity. The goal is to find sublinear time algorithms. An adaptation of the \local~model, the {\sc Slocal} model~\cite{ghaffari2017complexity} considers algorithms that are executed on nodes one after another. In particular, locally greedy problems are solved in $O(1)$ rounds in this model.

Bitton \etal~\cite{bitton2021fully} designed a self-stabilizing transformer for \local~problems. Their probabilistic transformer converts a given fault-free synchronous algorithm for \lcl~problems into a self-stabilizing synchronous algorithm for the same problem in anonymous networks.  The overheads of this transformation in terms of message complexity and average time complexity are upper bounded: the produced algorithms stabilize in time proportional to $\log(\alpha + \Delta)$ in expectation, where $\alpha$ is the number of faulty nodes.

Awerbuch \etal~\cite{awerbuch1989network} introduced the ruling set as a tool for decomposing the graph into small-diameter connected components. As for the seminal work,  the ruling set problems have been used as a  sub-routine function   to solve some other distributed problems (network decompositions~\cite{awerbuch1989network,barenboim2016locality}, colorings~\cite{panconesi1995local},   shortest paths~\cite{henzinger2019deterministic}).

The MIS problem has been extensively studied in the \local~model, \cite{ghaffari2016improved,rozhovn2020polylogarithmic,censor2020derandomizing} for instance and in the {\sc Congest} model~\cite{peleg2000distributed} (synchronous model where messages are $O(\log n)$ bits long).
In the \local~model, Barenboim \etal~\cite{barenboim2018locally} focus on systems with unique identifiers and gave a self-stabilizing algorithm producing an MIS within $O(\Delta + \log^{*} n)$ rounds.
Balliu \etal~\cite{balliu2021lower} prove that the previous algorithm~\cite{barenboim2018locally} is optimal for a wide range of parameters in the \local~model. In the {\sc Congest} model, Ghaffari \etal~\cite{ghaffari2021improved} prove that there exists a randomized distributed algorithm that computes a maximal independent set in $O(\log \Delta \cdot \log \log n + \log^{6} \log n)$ rounds with high probability.  Considering the problem $(\alpha,\beta)$-ruling set in a more general way,
Balliu \etal~\cite{balliu2022distributed} give some lower bound for computing a $(2, \beta)$-
ruling set in the \local~model:  any deterministic algorithm requires $\Omega \left( \min \quickset{  \frac{\log \Delta}{\beta \log \log \Delta}  , \log n } \right)$ rounds.

Up to our knowledge, no self-stabilizing algorithm has been designed for only computing $(k,k-1)$-ruling sets where $k>2$ under the Gouda daemon. 
Self-stabilizing algorithms for maximal independent set have been designed in various models (anonymous network~\cite{shukla1995observations,turau2006randomized,turau2019making} or not~\cite{goddard2003self,ikeda2002space,turau2007linear}).  Shukla~\etal~\cite{shukla1995observations} present the first self-stabilization algorithm designed for finding an MIS  for anonymous networks. Turau~\cite{turau2007linear} gives the best-known result with $O(n)$ moves under the distributed daemon. Recently, some works improved the results in the synchronous model. For non-anonymous networks, Hedetniemi \cite{hedetniemi2021self} designed a self-stabilization algorithm that stabilizes in $O(n)$ synchronous rounds. Moreover, for anonymous networks, Turau~\cite{turau2019making}~designs some randomized self-stabilizing algorithms for maximal independent set that stabilizes in $O(\log n)$ rounds w.h.p.
See the survey~\cite{guellati2010survey} for more details on MIS self-stabilizing algorithms.

\subsection{Model}
\label{sec:model}
A distributed system consists of a set of processes where two adjacent processes can communicate with each other. The communication relation is represented by a graph $G = (V,E)$ where $V$ is the set of the processes (we  call \motnouveau{node} any element of $V$ from now on) and $E$ represents the neighborhood relation between them, \textit{i.e.}, $uv \in E$ when $u$ and $v$ are adjacent nodes. 
The set of \motnouveau{neighbors} of a node $u$ is denoted by $N(u)$. 
We assume the system to be \motnouveau{anonymous}, meaning that a node has no identifier. Moreover, we consider undirected networks (i.e. $uv\in E\iff vu\in E$).

For communication, we consider the \motnouveau{shared memory model}:  the \motnouveau{local state} of each node corresponds to a set of \motnouveau{local variables}. 
A node can read its local variables and its neighbors'
but can only rewrite its local variables. A \motnouveau{configuration} is the value of the local states of all nodes in the system.
When $u$ is a node and $x$ is a local variable, the \motnouveau{$x$-value} of $u$ is the value $x_u$.
Each node executes the same algorithm that consists of a set of \motnouveau{rules}. Each rule is of the form \motnouveau{``\textbf{if} $\langle guard \rangle$ \textbf{then} $\langle command \rangle$"} and is parameterized by the node where it would be applied. Each rule also has a priority number.
The \motnouveau{guard} is a predicate over the variables of the current node and its neighbors. The \motnouveau{command} is a sequence of actions that may change the values of the node's variables (but not those of its neighbors).
A rule is \motnouveau{activable} in a configuration $C$ if its guard in $C$ is true. A process is \motnouveau{eligible} for the rule $\mathcal{R}$ in a configuration $C$ if its rule $\mathcal{R}$ is activable and no rule of lower priority number is activable for that node in $C$. We say in that case that the process is \motnouveau{activable} in $C$.  An \motnouveau{execution} is an alternate sequence of configurations and actions  $\Exe = C_0,A_0,\ldots,C_i,A_i, \ldots$, such that $\forall i\in \mathbb{N}^*$, configuration $C_{i+1}$ is obtained by executing the command of at least one rule
that is activable  in configuration $C_i$. 
More precisely, the set of actions  $A_i$ is the non-empty set of activable processes in $C_i$ such that their activable rules have been executed to reach $C_{i+1}$. 

The goal of a self-stabilizing algorithm is to be robust to perturbations.
An initial configuration cannot follow any restriction, and failures can occur, changing the state of some of the nodes.
A self-stabilizing algorithm must be able to recover and reach a correct general output from any configuration.

In a distributed system, multiple nodes can be active simultaneously, meaning they are in a state where they can make a computation.  
The definition of a self-stabilizing algorithm is centred around the notion of \motnouveau{daemon}. A  \motnouveau{daemon} captures which set of activable rules are chosen by some scheduler during the execution. 
See~\cite{dubois2011taxonomy} for  a taxonomy.  In our work, we consider the \motnouveau{Gouda} daemon.

\begin{definition}\cite{dubois2011taxonomy,gouda2001theory} We say that an execution $  \Exe = C_0\rightarrow C_1 \rightarrow C_2\dots$ is \motnouveau{under the  Gouda deamon} if: for any configurations $C$ and $C'$ such that $C \rightarrow C'$ can be executed, if $C$ appears infinitely often in $\Exe$, then $C'$ also appears infinitely often in $\Exe$. 

 

\end{definition}



An algorithm is \motnouveau{self-stabilizing} for a given specification (i.e. a set of restrictions over the configurations) under some daemon if there exists a subset $\cal L$ of the set of all configurations, called the \motnouveau{legitimate configurations}, such that:
\begin{itemize} 
\item Any configuration in $\cal L$ verifies the specification, and any execution under the said daemon starting in $\cal L$ stays in $\cal L$. (\motnouveau{correctness}).
\item Any execution under the said daemon eventually reaches a configuration in $\cal L$ (\motnouveau{convergence}).
\end{itemize}
The set $\cal L$ is called the set of \motnouveau{legitimate configurations}.


%% file: algo.tex
\begin{figure} 
\setlength\abovecaptionskip{0pt}
{\footnotesize
\begin{tabbing}
aa\=aaa\=aaa\=aaaaaaaa\=\kill
\textbf{-------------- Attributes of the nodes}\\[0.5em]

\textbf{d$_u \in \segment{0}{k-1}$} 
~\\

\textbf{err$_u \in \quickset{0,1}$}
~\\

For every $i \in \segment{1}{\floor{\frac{k}{2}}-1 }$ : \textbf{c$_{i,u} \in \mathbb{Z}/4 \mathbb{Z}$} 
 and 
\textbf{b$_{i,u} \in \quickset{\uparrow,\downarrow}$} 
~\\

\textbf{-------------- Predicates}\\[0.5em]

well\_defined$(u) \equiv err_u = 0 \wedge 
\forall v \in N(u), |d_u -d_v| \leq 1 \wedge (d_u>0 \Rightarrow (\exists v \in N(u), d_v = d_u-1))$\\ 
leader\_down$(u) \equiv d_u=0 \Rightarrow \forall i \in \segment{1}{\ksurdeux -1}, b_{i,u}=\downarrow$\\ branch\_coherence\_up$(u,i) \equiv$ \\
\hspace{0.5cm} $\forall v\in N(u),
d_v = d_u-1 \Rightarrow (b_{i,u}, b_{i,v}, c_{i,v}) \in \quickset{(\uparrow,\uparrow,c_{i,u}),(\uparrow,\downarrow,c_{i,u}),(\uparrow,\downarrow,c_{i,u}+1),(\downarrow,\downarrow,c_{i,u})}$ \\

branch\_coherence\_down$(u,i) \equiv$ \\
\hspace{0.5cm}$\forall v\in N(u),
d_v = d_u+1 \Rightarrow (b_{i,u}, b_{i,v}, c_{i,v}) \in \quickset{(\uparrow,\uparrow,c_{i,u}),(\downarrow, \uparrow,c_{i,u}),(\downarrow, \uparrow,c_{i,u}-1),(\downarrow,\downarrow,c_{i,u})}$ \\


branch\_coherence$(u) \equiv$ 
$d_u \geq \ksurdeux \vee \big(branch\_coherence\_up(u,d_u)
\; \; \wedge $ \\
\hspace{1.8cm}$\forall i \in \segment{d_u+1}{\floor{\frac{k}{2}}-1}, branch\_coherence\_up(u,i)\wedge branch\_coherence\_down(u,i)\big)
$~\\


\textbf{-------------- Rules}\\

\textbf{Incr Leader::} (priority 2) \\
\> \textbf{if} $\text{ well\_defined}(u)\wedge (d_u=0)\wedge(\exists i\in \segment{1}{\floor{\frac{k}{2}}-1}, \forall v \in N(u), d_v = 1 \wedge c_{i,u}-c_{i,v}=0)$\\ 
\> \textbf{then} For all such $i$, $c_{i,u}:= c_{i,u}+1$ \\ 
~\\
\textbf{Sync 1 down::} (priority 2) \\
\> \textbf{if} $\text{ well\_defined}(u) \wedge \exists !v \in N(u), \exists i\in \segment{1}{\floor{\frac{k}{2}}-1}, d_u = 1 \wedge d_v = 0 \wedge c_{i,u}=c_{i,v}-1 \wedge b_{i,u} = \uparrow )$\\
\> \textbf{then} For all such $i$, $c_{i,u}:= c_{i,v}$ ; $b_{i,u} := \downarrow$ \\
~\\

\textbf{Sync 2+ down::} (priority 2) \\

\> \textbf{if} $\text{ well\_defined}(u) \wedge 1 < d_u < \floor{\frac{k}{2}}$\\
$\wedge (\exists i\in \segment{d_u}{\floor{\frac{k}{2}}-1},b_{i,u} = \uparrow \wedge \forall v \in N(u),d_v = d_u-1 \Rightarrow (c_{i,u}=c_{i,v}-1 \wedge b_{i,v} = \downarrow ))$\\
\> \textbf{then} For all such $i$, $c_{i,u}:= c_{i,v}$ ; $b_{i,u} := \downarrow$ \\
~\\

\textbf{Sync 1+ up::} (priority 2) \\

\> \textbf{if} $\text{ well\_defined}(u) \wedge 0 < d_u < \floor{\frac{k}{2}}$\\
$\wedge (\exists i\in \segment{d_u+1}{\floor{\frac{k}{2}}-1}, b_{i,u} = \downarrow  \wedge \forall v \in N(u),d_v = d_u+1 \Rightarrow (c_{i,u}=c_{i,v} \wedge b_{i,v} = \uparrow ))$\\
\> \textbf{then} For all such $i$, $b_{i,u} := \uparrow$ \\
~\\

\textbf{Sync end-of-chain::} (priority 2) \\

\> \textbf{if} $\text{ well\_defined}(u) \wedge 0 < d_u < \floor{\frac{k}{2}} \wedge \forall v \in N(u),d_v = d_u-1 \Rightarrow (c_{d_u,u}=c_{d_u,v}-1 \wedge b_{i,v} = \downarrow ))$\\
\> \textbf{then} $b_{d_u,u} := \uparrow$ ; $c_{d_u,u} := c_{i,v}$\\
~\\

%





\textbf{Update distance ::} (priority 0) \\
\> \textbf{if} $(d_u\not=0) \wedge d_u \neq \min(\min\quickset{d_v|v\in N(u)}+1,\KAmoinsUn)$\\

\> \textbf{then} $d_u := \min(\min\quickset{d_v|v\in N(u)}+1,\KAmoinsUn)$\\
\hspace{1.0cm} If $d_u < \floor{\frac{k}{2}}$ : Let $v:= \text{choose}(\quickset{w \in N(u) | d_w=d_u-1})$ \\
\hspace{2.9cm} For each $i \in \segment{d_u}{\ksurdeux-1}, c_{i,u}:= c_{i,v}$ ; $b_{i,u} := b_{i,v}$\\

~\\

\textbf{Become Leader ::} (priority 2) \\
\> \textbf{if} $err_u=0 \wedge (d_u=\KAmoinsUn) \wedge \forall v \in N(u), d_v =\KAmoinsUn$\\

\> \textbf{then} $d_u := 0$, 
For each $i \in \segment{1}{\floor{\frac{k}{2}}-1}, c_{i,u}:=0,b_{i,u} := \downarrow$ \\
~\\

\textbf{Leader down ::} (priority 1) \\
\> \textbf{if} $well\_defined(u) \wedge d_u=0 \wedge \exists i \in \segment{1}{\floor{\frac{k}{2}}-1}, b_{i,u}=\uparrow$  \textbf{then} For each $i\in \segment{1}{\floor{\frac{k}{2}}-1}$, $b_{i,u} := \downarrow$\\
~\\

\textbf{Two Heads::} (priority 1) \\
\> \textbf{if} $err_u=0 \wedge \exists v,v' \in (N(u)\cup \quickset{u})^2, v \not=v' \wedge d_v = d_{v'} = 0)$  \hspace{1cm} \textbf{then} $err_u := 1$ \\
~\\

\textbf{Branch incoherence::} (priority 1) \\
\> \textbf{if}
$err_u=0 \wedge \neg branch\_coherence(u)$ \hspace{4cm} \textbf{then} $err_u := 1$ \\
~\\
\textbf{Error Spread ::} (priority 2) \\
\> \textbf{if} $err_u=0 \wedge (d_u \leq  \floor{\frac{\KA}{2}} -1) \wedge (\exists v \in N(u), err_v =1 \wedge d_u < d_v)$  \hspace{0.5cm} \textbf{then} $err_u := 1$\\
~\\



\textbf{Reset Error ::} (priority 2) \\
\> \textbf{if} $(err_u=1) \wedge ([d_u > \floor{\frac{k}{2}}] \vee [\forall v \in N(u), d_v \geq d_u \vee err_v=1])$\\
\> \textbf{then} $err_u:=0$,  
If $d_u=0$, $d_u := 1$, 
For each $i$, $c_{i,u} := 0$, $b_{i,u} := \uparrow$
\end{tabbing}
}
\caption{Algorithm for the $(k,k-1)$-Ruling Set}

\end{figure}

%% file: Explanation.tex
\section[Self-Stabilizing Algorithm for Computing a (k, k-1)-Ruling 
Set]{Self-Stabilizing Algorithm for Computing a $(k, k-1)$-Ruling 
Set}\label{Sec:RulingSet}

\subsection{General Overview}

As we want to compute a $(k,k-1)$-ruling set, a node needs to detect when it is currently ``too far'' from the nodes pretending to be in the ruling set. When $k=2$, 
a $(2,1)$-ruling set is an MIS, and some
 self-stabilization algorithms are designed for finding an MIS  \cite{shukla1995observations,turau2006randomized,turau2019making}.
For the remaining of the document, we assume $k>2$.

To this aim, the local variable $d$  represents the distance at which the node thinks it is from the ruling set.
In particular, a $d$-value of $0$ indicates that a node is (or thinks it is) in the ruling set, and we denote by $S(C)$ the set of those nodes in a given configuration $C$.
Any other value of $d_u$ represents the distance to $S(C)$ (the minimum between $k-1$ and the said distance). The rule \textbf{Update distance} has the highest priority. Its goal is to ensure that each node eventually gets its distance to $S(C)$ accurately. 
When a node $u$ has its local variable $d_u$ equal to $k-1$ and is surrounded by nodes of $d$-value $k-1$, it ``knows'' that it is far enough from $S(C)$ to be added to it. Node $u$ can then execute rule \textbf{Become Leader} to do so. Update of $d$-values will then spread from the new member of $S(C)$ through the execution of rule \textbf{Update distance}.

The way to insert new nodes into $S(C)$ cannot avoid the fact that two new members of $S(C)$ may be too close.
A way to detect those problems is needed to guarantee that we will not let those nodes in $S(C)$.

 If they are close enough (distance $2$ or less), it can be directly detected by a node (either a common neighbor if they are at distance $2$ or one of them if they are at distance $1$). The rule \textbf{Two Heads} is here to detect this. 



No node can detect this problem when problematic nodes are too far away.
To remedy this, each node maintains a synchronized clock system around each node of $S(C)$ by executing the \motnouveau{stationary rules}. For this reason, 
we split the set of rules into two groups:
\begin{itemize}
\item The \motnouveau{stationary} rules are the rules  \textbf{Incr Leader}, \textbf{Sync 1 down},  \textbf{Sync 2+ down}, \textbf{Sync 1+ up}, and \textbf{Sync end-of-chain};
 \item The \motnouveau{convergence} rules are the rules \textbf{Remote Collision}, \textbf{Two Heads}, \textbf{Branch Incoherence}, \textbf{Update Distance}, \textbf{Become Leader}, \textbf{Error Spread}, \textbf{Reset Error}, and \textbf{Leader down}.
\end{itemize}

We say that a node in $S(C)$  is the \motnouveau{leader} of the nodes under its influence, corresponding to the nodes in its ball at distance $\ksurdeux$. Assuming $d$-value has already been spread, the clock of index $i$ of nodes that gave the same leader will always be either equal or out-of-sync by $1$. Thus, a node detects that two nodes in $S(C)$ are too close when it sees in its neighbourhood two nodes with clocks out-of-sync by $2$. It will raise an error when activated by executing rule \textbf{Remote Collision}. 
The error is then propagated toward the problematic members of $S(C)$ by rule \textbf{Error Spread}.   

In both previous cases, the problematic nodes of $S(C)$ end up having $err$-value $1$, which makes them leave $S(C)$ by executing rule \textbf{Reset Error}. Afterwards, rule \textbf{Update distance} will, over time, update the $d$-values of the nodes at distance up to $k$ to that node. 

The goal of our algorithm is to ensure that we reach locally a configuration from which, when a node is inserted in $S(C)$, and no node gets added at distance at most $k-1$ away, it remains in $S(C)$ forever. Note that when it is executed, rule \textbf{Update distance} setup the clock values and arrows (variables $c$ and $b$) so that the newly updated node is synchronized to its ``parent'' (the node it takes as a reference to update its $d$-value).




The target configuration is \textbf{not} a \motnouveau{stable} configuration, and   from it, all the nodes  can only   execute      \motnouveau{stationary rules}. In this configuration, $S(C)$ is guaranteed to be a $(k,k-1)$-ruling set of the underlying graph. Note that the predicate $well\_defined$ appears in the guard of every stationary rule. The predicate guarantees that the considered node neither is in error-detection mode nor has some incorrect $d$-values in its neighbourhood before executing any clock-related rule.

\subsection{The Clock System}

Now, we describe the clock system that detects two leader nodes in $S(C)$ at a distance less than $k$. The leaders are the nodes that update the clock value $c_i$ and propagate it to its ``children'' and so on.
%
%
%
For a given clock index $i$, when every neighbour of a leader $s$ has the same clock $c_i$ and their corresponding arrow $b_i$ pointed up, node $s$  increments its clock value by $1$ by executing rule \textbf{Incr Leader}.

After that, the clock value is propagated downward (toward nodes of greater $d$-value) using rules \textbf{Sync 1 down} and \textbf{Sync 2+ down}. Note that it is performed locally by layers: one node of a given $d$-value cannot update its clock value and arrow before every neighbour with a smaller $d$-value does so. This is necessary to guarantee the global synchronization of the clock.

There are two ways for the propagation of $(c_i,b_i)$ to reach the limit of the area it should spread in:  either it has reached nodes with $d$-value $i$, or there is no node having a greater $d$-value to spread the clock further.
\begin{itemize}
\item In the first case, rule \textbf{Sync end-of-chain} flips the arrow $b_i$.
\item In the second case, the nodes execute rule \textbf{Sync 1+ up} to flip $b_i$.
\end{itemize}
In both cases, it allows rule \textbf{Sync 1+ up} to propagate upward (toward smaller $d$-values) with the $b_i$-value switching to $\uparrow$ from the nodes to their parents. Note that it is done locally by layers: one node of a given $d$-value may not update its clock value and arrow before every neighbour with a greater $d$-value has done so.

When the propagation reaches the neighbors of $s$, node $s$ ``detects'' that its current clock value has been successfully propagated, and it will execute rule \textbf{Incr Leader} to increase it.

The point of this clock system is that two nodes under the same leader cannot have clock values out-of-sync by $2$, but two nodes that have different leaders may. It allows them to detect a ``collision'' (\textit{i.e.} two nodes of $S(C)$ too close from each other) when the $d$-values of two such nodes are smaller than $\ksurdeux$. Observe that the clock of index $i$ is only reliable for detecting collision between nodes of $S(C)$  at distance $2i$ or $2i+1$ from each other. For smaller distances, this clock may be forcefully synchronized between two nodes of $S(C)$ by layer-by-layer updating, and for greater distances, no node may detect an out-of-sync from it. This process differs from the PIF mechanism~\cite{cournier2001selfPIF}:  we need to run one clock for each layer, as a clock of higher layer will be synchronized for the two conflicting leaders because of further nodes. Thus, we have $\ksurdeux-1$ parallel clock systems to capture every possible distance  of collision. 

The Gouda daemon ensures that if two nodes of $S(C)$ are too close, this will only be the case for a while. The clock system will eventually detect it and propagate an error.


\subsection{Handling Initial and Perturbed Configurations}

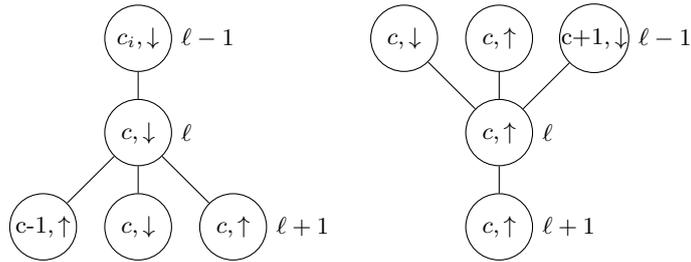
\begin{figure}
\begin{center}
\input{branch_coherence.tikz}
\end{center}
\caption{Branch coherence condition. The couples $(c,\downarrow)$ or $(c,\uparrow)$ represent the local variables $(c_i,b_i)$ of the nodes. The value on the right of the node represents its distance to the leader node (i.e. its $d$-value). The central node in both figures is the reference, and the other nodes represent the possible couples for its neighbors with different $d$-value.\\
}
\label{fig:branchcoherence}
\end{figure}

Rules \textbf{Leader Down} and \textbf{Branch Incoherence} are only executed to solve problems coming from the initial configuration or after a perturbation has occurred. Rule \textbf{Leader Down} is executed when a leader has some of its arrows $b_i$ in the wrong direction. 
Rule \textbf{Branch Incoherence} is executed when some ``impossible'' patterns are produced in the clock systems due to wrong clock values and arrows in the initial state. Standard patterns are shown in Figure~\ref{fig:branchcoherence}. Any other pattern will make an activated node to execute rule \textbf{Branch Incoherence}.

%% file: branch_coherence.tikz
\begin{tikzpicture}

\tikzstyle{circlenode}=[draw,circle,minimum size=70pt,inner sep=0pt]
\tikzstyle{whitenode}=[draw,circle,fill=white,minimum size=25pt,inner sep=0pt]

\draw (0,1.25) node[whitenode] (a1) [label=0:{\small $\ell-1$}]  {\small $c_i,\downarrow$};

\draw (0,0) node[whitenode] (a0) [label=0:{\small $\ell$}]{\small $c,\downarrow$};
https://forms.gle/GmSs6odZRKRp3VX98
\draw (-1.25,-1.25) node[whitenode] (a2) {\small $\scalebox{1}[1.0]{c-1},\uparrow$};
\draw (0,-1.25) node[whitenode] (a3) {\small $c,\downarrow$};
\draw (1.25,-1.25) node[whitenode] (a4) [label=0:{\small $\ell+1$}]  {\small $c,\uparrow$};

\draw (a0) edge node {} (a1);
\draw (a0) edge node {} (a2);
\draw (a0) edge node {} (a3);
\draw (a0) edge node {} (a4);

\draw (3.5,1.25) node[whitenode] (b2) {\small $c,\downarrow$};
\draw (4.75,1.25) node[whitenode] (b3) {\small $c,\uparrow$};
\draw (6,1.25) node[whitenode] (b4) [label=0:{\small $\ell-1$}]  {\small $\scalebox{1}[1.0]{c+1},\downarrow$};

\draw (4.75,0) node[whitenode] (b0) [label=0:{\small $\ell$}]{\small $c,\uparrow$};

\draw (4.75,-1.25) node[whitenode] (b1) [label=0:{\small $\ell+1$}]  {\small $c,\uparrow$};

\draw (b0) edge node {} (b1);
\draw (b0) edge node {} (b2);
\draw (b0) edge node {} (b3);
\draw (b0) edge node {} (b4);

\end{tikzpicture}

%% file: proof.tex
\section{Proof of the Algorithm}\label{sec:proof}

\begin{articlecourt}
\articlecourt{The file contains two versions of the paper: the first version without the proofs, and the second complete version that contains all the proofs.}
\end{articlecourt}

\subsection{Stability of Legitimate Configurations}

The ruling set algorithm presented in this section uses the state model. It constructs the set of vertices whose $d$-value is $0$. We will prove that this set is a ruling set in legitimate configurations. Formally, we require the following specification for the legitimate configurations:

\begin{definition}\label{def:legitimate}
 Let $S(C)$  be the set of nodes $s$ such that   $d_{s} =0$ in a given configuration $C$.
Configuration $C$ is said to be \motnouveau{legitimate} if:
\begin{enumerate}
\item for any $u$ we have $well\_defined(u)$, $leader\_down(u)$ and $branch\_coherence(u)$ hold;
\item for any two distinct nodes $u$ and $v$ of $S(C)$, we have $dist(u,v) \geq k$.
\end{enumerate}
\end{definition}

\begin{theorem} \label{th:stability}
The set of legitimate configurations is closed.
 Moreover, all the $d$-values do not change from a legitimate configuration $C$.
\end{theorem}

Thanks to Theorem~\ref{th:stability}, we know that, from a legitimate configuration, we keep the same set of leaders $S(C)$, which forms a $(k,k-1)$-ruling set. Hence, under the Gouda daemon, the set of leaders will eventually be a stable $(k,k-1)$-ruling set.

The goal of the following lemmas will be to prove Theorem~\ref{th:stability}.
Lemma~\ref{lem:legitimate:prop} ensures that $S(C)$ forms a ruling set when the values of all the local   variables  are correct.
\begin{articlelong}
\begin{lemma} \label{lem:legitimate:prop}
Let $C$ be a legitimate configuration.
\begin{itemize}
 \item  For any node $u$, $d_{u} = dist(u,S(C))$;
 \item $S(C)$ is a $(k, k-1)$-ruling set of the underlying graph.
\end{itemize}
\end{lemma}
\end{articlelong}

\begin{articlecourt}
\begin{lemma} \label{lem:legitimate:prop}
Let $C$ be a legitimate configuration. For any node $u$, $d_{u} = dist(u,S(C))$, and $S(C)$ is a $(k, k-1)$-ruling set of the underlying graph.
\end{lemma} 
\end{articlecourt}

\begin{articlelong}\begin{proof} 

By the definition of predicate $well\_defined(u)$, if $d_u>0$, there exists some $v\in N(v)$ such that $d_v = d_u-1$. Let us prove by induction on $i<k$ that ``For any node $u$, $d_u\le i\Leftrightarrow dist(u,S(C))=d_u$'':

\begin{itemize}
    \item If $d_u=0$, we have $u\in S(C)$, and $dist(u,S(C))=dist(u,u)=0$. Conversely, if $dist(u,S(C))=0$, then   $u\in S(C)$  and  $d_u=0$ by definition of $S(C)$.
    \item Let us assume that it is true for $i<k-1$. 
    Let $u$ be a node such that $d_u=i+1$.  Since predicate $well\_defined(u)$ is satisfied in $C$, there exists a node $v\in N(u)$ such that $d_v = d_u-1=i$. Using induction hypothesis, $dist(v,S(C))=i$, \motnouveau{i.e.} there exists node $s$ in $S(C)$ such that $dist(s,v)=i$. Hence, $dist(s,u)\le i+1$.
    Using induction hypothesis, $dist(u,S(C))\leq i$ would imply $d_u=dist(u,S(C)) \leq i$, thus by contraposition $dist(u,S(C))> i$. 
    Hence $dist(u,S(c))=i+1$.

    Conversely, Let us assume that $dist(u,S(C))=i+1$. By using a shortest path from $u$ to $S(C)$, we can get a node $v\in N(u)$ such that $dist(v,S(C))=i$. Hence, by induction, $d_v=i$. By the predicate $well\_defined(u)$, we have $i-1\le d_u\le i+1$, since $|d_u -d_v| \leq 1$. As $dist(u,S(C))=i+1\not\in \segment{0}{i}$, by induction hypothesis we get $d_u > i$. Hence $d_u=i+1$.
\end{itemize}
By induction, we proved the first item of the lemma.

Let us prove that $S(C)$ is a $(k, k-1)$-ruling set. Since every node $u$ has the value of local variable $d_u$  less than $k-1$, the first item of this lemma states that $dist(u,S(C))\le k-1$. Hence, we only need that for any pair $(u,v)\in S(C)^2$ of distinct nodes, $dist(u,v)\ge k$. The latter fact is true by definition of a legitimate configuration. Thus, the lemma holds.
\end{proof}\end{articlelong}

\begin{figure}
\begin{center}
\begin{tikzpicture}

\tikzstyle{circlenode}=[draw,circle,minimum size=70pt,inner sep=0pt]
\tikzstyle{whitenode}=[draw,circle,fill=white,minimum size=25pt,inner sep=0pt]
 
\draw (0,0) node[whitenode] (a0) [label=-90:{\small$c,\downarrow$}]  [label=90:{\small $d_s=0$}]{\small$s$};
\draw (1.75,0) node[whitenode] (a1) [label=-90:{\small$c,\downarrow$}]  {\small$s_1$};
\draw (3.5,0) node[whitenode] (a2) [label=-90:{\small$c,\downarrow$}]  {\small$s_a$};
\draw (5.25,0) node[whitenode] (a3) [label=-90:{\small$c',\uparrow$}]  {\small$s_{a+1}$};
\draw (7,0) node[whitenode] (a4) [label=-90:{\small$c',\uparrow$}]  {\small$s_{d_u-1}$};
\draw (8.75,0) node[whitenode] (a5) [label=90:{\small $d_u$}] {\small$u$};

\draw (a0) edge node {} (a1);
\draw (a2) edge node {} (a3);
\draw (a4) edge node {} (a5);
\draw (a1) edge [dotted]node {} (a2);
\draw (a3) edge [dotted]node {} (a4);

\end{tikzpicture}
\caption{Node $s$ propagates its clock value  along a shortest path from $s$ to $u$ where $c'\in\{c,c-1\}$.} \label{Fig:shortestpath}
\end{center}
\end{figure}
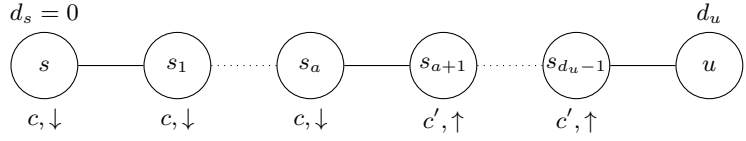

Now we  focus on the clock system.  We  prove the following property on the ruling set to run the clock system. 

\begin{lemma}\label{lem:legitimate:closenodes}
Let $C$  be a legitimate configuration and $s$  be a node in $S(C)$. For every node $u$,  $dist(u,s) \leq \ksurdeux$ implies that $ d_u = dist(u,s)$.
\end{lemma}
\begin{articlelong}\begin{proof}
Suppose node $u$ is such that $dist(u,s) \leq \ksurdeux $.

Let us take $s' \in S(C)$ such that $d(u,S(C)) = d(u,s')$, we then have $dist(u,s) \leq dist(u,s')$. Then by triangular inequality we have $dist(s,s') \le dist(s,u) + dist(u,s') \leq 2\, dist(s,u) \leq  2\ksurdeux \leq k-1$. As $S(C)$ is a $(k,k-1)$-ruling set from Lemma~\ref{lem:legitimate:prop}, this means that $s=s'$.
\end{proof}\end{articlelong}

This property  
allows us to deduce that a node $u$  such that $dist(u,s) \leq \ksurdeux$   has only one node $s$ of $S(C)$ in its ball at distance $\ksurdeux$. Thus, all the nodes in $\mathcal{B}(s,\ksurdeux-1)$ must be synchronized with $s$.
 We explain how the values representing the clock of the local variable of nodes with $d$-value smaller than $\ksurdeux$ are spread from their leader. Figure~\ref{Fig:shortestpath} illustrates how the pairs $(c_i,b_i)$ go from nodes in $S(C)$.
\begin{lemma} \label{lem:legitimate:clock}
Let $C$ be a legitimate configuration and $s$ a node in $S(C)$.
For every node $u$ such that $dist(u,s) \leq \ksurdeux-1$, every shortest path $(s_0,s_1,\cdots,s_{d_u})$ from $s$ to $u$ satisfies the  following property in $C$:\\
For every clock index $i \in \segment{d_u+1}{\ksurdeux-1}$,
there exists some integer $a\in \segment{0}{d_{u}}$ such that:
\begin{enumerate}
\item $\forall \ell \in \segment{0}{a}$,  $(b_{i,s_\ell},c_{i,s_\ell}) = (\downarrow,c_{i,s}) $;
\item $\exists c'\in\{c_{i,s}-1,c_{i,s}\}, \forall \ell \in \segment{a+1}{d_u}$,  $(b_{i,s_\ell},c_{i,s_\ell})=(\uparrow,c')$.
\end{enumerate}
\end{lemma} 
\begin{articlelong}\begin{proof}
For a given integer $\alpha \in \segment{0}{\ksurdeux-1}$ we denote by $\mathcal{H}(\alpha)$ the property of the lemma for every node whose distance to $s$ is $\alpha$. 


For $\alpha = 0$, the only path to consider only contains $s$ itself. Consider any clock index $i \in \segment{1}{\ksurdeux-1}$. The only possible value for $a$ is $0$.
\begin{enumerate}
    \item By definition of the legitimate configuration, $b_{i,s}= \downarrow$. Thus $(b_{i,s},c_{i,s}) = (\downarrow,c_{i,s}) $, and Point~$1$ of property $\mathcal{H}(0)$ holds.
    \item Point~$2$ of property $\mathcal{H}(0)$ holds trivially, since $a+1 > d_u =0 $.
\end{enumerate}
Thus $\mathcal{H}(0)$ holds.

We now suppose $\mathcal{H}(\alpha)$ true for any $\alpha < \delta$ for some $\delta \leq \ksurdeux-1$.
Let then $P= (s=s_0,s_1,\cdots,s_{\delta}=u)$ be a shortest path from node $s$ to some node $u$ such that $dist(s,u)=\delta$.
Since $C$ is legitimate, $d(u,S(C)) = d(u,s)$ by Lemma~\ref{lem:legitimate:closenodes} since $\delta \leq \ksurdeux$.
Lemma~\ref{lem:legitimate:prop} gives then $d_u=dist(s,u)=\delta$.

By definition of a shortest path, $s_{\delta-1}$   is at distance $dist(s,u)-1= \delta-1$ from $s$. By the same argument as for $u$, Lemma~\ref{lem:legitimate:prop} gives $d_{s_{\delta-1}}=\delta-1$, thus $d_{s_{\delta-1}} < \delta$.
Let also $i$ be an integer  in $\segment{d_u+1}{\ksurdeux-1}$ be a clock index.
Since  $dist(s,s_{\delta-1}) < \delta$, we  apply the induction hypothesis -- the shortest path $(s_0,s_1,\cdots,s_{\delta-1})$ from  $s$ to $s_{\delta-1}$
 satisfies that there exists some $a\in[0,\delta-1]$ such that:
\begin{enumerate}
\item $\forall \ell \in \segment{0}{a}$,  $(b_{i,s_\ell},c_{i,s_\ell}) = (\downarrow,c_{i,s}) $;
\item $\exists c'\in\{c_{i,s}-1,c_{i,s}\}, \forall \ell \in \segment{a+1}{\delta-1}, (b_{i,s_\ell},c_{i,s_\ell})=(\uparrow,c')$.
\end{enumerate}

We treat the cases $a=\delta -1$  and $a<\delta -1$ separately, using the same argument that predicate $branch\_coherence\_down(s_{\delta-1},i)$  is true (as $\delta-1 < \delta \leq i <\frac{k}2$, and $C$ is legitimate).

Suppose that $a<\delta -1$. We have $(b_{i,s_{\delta-1}},c_{i,s_{\delta-1}})=(\uparrow,c')$ for some $c'\in\{c_{i,s}-1,c_{i,s}\}$.
 Since predicate $ branch\_coherence\_down(s_{\delta-1},i)$ is true,   the fact that $b_{i,s_{\delta-1}}=\uparrow$  implies  that  $ (b_{i,s_{\delta}},c_{i,s_{\delta}})=(\uparrow,c')$.
Thus, $P= (s_0,s_1,\cdots,s_{\delta})$ respects the property of the lemma.

Suppose that $a=\delta -1$. Using induction hypothesis, we have  $\forall \ell \in \segment{0}{\delta-1}, (b_{i,s_{\ell}},c_{i,s_{\ell}}) = (\downarrow,c_{i,s})$. Since  $branch\_coherence\_down(s_{\delta-1},i)$  is valid, we have three possibilities:

\begin{itemize}
\item If $(b_{i,u},c_{i,u}) = (\downarrow,c_{i,s_{\delta -1}}) $, we have $(b_{i,u},c_{i,u}) = (\downarrow,c_{i,s})$.
Then $P= (s_0,s_1,\cdots,s_{\delta})$ respects the wanted property for $a=\delta$. Note that the second part of the property is trivially true with this value of $a$ for any value $c'$, for the ``for all'' quantifier acts on the empty set.

\item If $(b_{i,u},c_{i,u}) = (\uparrow,c_{i,s_{\delta -1}}) $, we have $(b_{i,u},c_{i,u}) = (\uparrow,c_{i,s})$. 
Then, $P= (s_0,s_1,\cdots,s_{\delta})$ respects the wanted property for $a=\delta-1$, taking $c'=c_{i,s}$ for the second part.

\item Else, we must have $(b_{i,u},c_{i,u}) = (\uparrow,c_{i,s_{\delta -1}}-1) $, and we have $(b_{i,u},c_{i,u}) = (\uparrow,c_{i,s}) $.
Then, $P= (s_0,s_1,\cdots,s_{\delta})$ respects the wanted property for $a=\delta-1$, taking $c'=c_{i,s}-1$ for the second part.
\end{itemize}


Thus, every shortest path $(s_0,s_1,\cdots,s_{\delta})$ from $s$ to every such node $u$ satisfies the property for every possible clock index $i \in \segment{d_u+1}{\ksurdeux-1}$, and $\mathcal{H}(\delta)$ holds.

Thus, by induction, the lemma holds.

\end{proof}\end{articlelong}
Lemma~\ref{lem:legitimate:rule} proves that only rules to update clocks are executed from  legitimate configurations:

\begin{lemma} \label{lem:legitimate:rule}
Let $C$ be a  legitimate configuration. Let $u$ be a node.
 Node $u$ only executes stationary rules from $C$.
\end{lemma} 
\begin{articlelong}\begin{proof}



Let $C$ be a legitimate configuration. Since every node satisfies predicates $well\_defined$, $branch\_coherence$ and $leader\_down$, none has its $err$-value equals to $1$, and rules $\textbf{Leader Down}$, $\textbf{Error Spread}$, $\textbf{Reset Error}$ and  $\textbf{Branch incoherence}$ cannot be executed.

From Lemma~\ref{lem:legitimate:prop}, $S(C)$ is a $(k, k-1)$-ruling set of the underlying graph: each element in $S(C)$ is at distance at least $k$ from another. 
 Thus, no two neighbouring nodes with their local variables $d$ equal to $0$ and rule $\textbf{Two Heads}$ cannot be executed in $C$. 
 
Let $u$ be a node. The first point of Lemma~\ref{lem:legitimate:prop} implies that
$d_{u}=\min \{dist(u,s) \mid  s \in S(C)\}$.  Observe that, by definition of local variable  $d_u$, we have $0\leq d_u\le k-1$.

From predicate $well\_defined$, we know that if $d_u>0$, $u$ has a neighbor $v$ such that $d_v=d_u-1$, and that any neighbor $w$ of $u$ are such that $d_u-1\le d_w\le d_u+1$.
Hence, $d_u = \min(\min\quickset{d_v \mid v\in N(u)}+1,\KAmoinsUn)$. We deduce that rule $\textbf{Update distance}$ cannot be executed in $C$.

Observe that, if $d_{u}=k-1$, $u$ has at least a neighbor $v$ is such that  $d_{v}=d_u -1=k-2$. This implies that no node can activate rule  \textbf{Become Leader}.




 

Let $u$ be a node that executes rule \textbf{Remote Collision}. We have $d_u \leq \frac{k-1}{2}$,  the existence of two nodes $v$ and $v'$ in $N(u) \cup \{u\}$ with  $d_{v}=d_{v'}$, and $|c_{d_v,v} - c_{d_v,v'}|=2$ (as local variable $c_{d_v}$ is defined, we have $d_v=d_{v'}\le\ksurdeux-1$). It means that there exist two nodes  $s$, $s'$ in $S(C)$ such that $dist(v,s)=d_v$ and $dist(u,s')=d_u\le\frac{k-1}{2}$. As $v$ is in the closed neighborhood of $u$, we have:
$$dist(s,s') \leq dist(s,v) + 1+dist(u,s') \leq d_v+1+d_u \leq (\ksurdeux-1)+1+\frac{k-1}{2}<2\frac{k}2$$

As $S(C)$ is a $(k,k-1)$-ruling set, we get that $s=s'$. By the same reasoning, we get that $v'$ is also at distance $d_{v'}$ to $s$.



 

By applying Lemma~\ref{lem:legitimate:clock} for both $v$ and $v'$, we get that $c_{d_v,v},c_{d_{v'},v'}$ are in  $\{c_{d_v,s}-1,c_{d_v,s}\}$. Thus it implies that  $|c_{d_v,v} - c_{d_v,v'}| \leq 1$, which contradicts the execution of  rule \textbf{Remote Collision}.

This concludes the proof.

 \end{proof}\end{articlelong}

Once the execution reaches a legitimate configuration $C$, we have proved that only 
stationary rules can be executed. The goal is to use that result and the previous lemmas to prove that  only legitimate configurations can be reached from $C$. This result will lead to the proof of Theorem~\ref{th:stability}.

\begin{articlelong}\begin{proof}{of Theorem \ref{th:stability}.}
Let $C$  and $C'$  be two   configurations such that  $C$ is legitimate and  $C\to C'$. 
By enumerating cases, we prove that $C'$ is also legitimate.

Based on the stationary rules, if $d_u=0$, for any $i<\floor{\frac k2}$, we always have $b_{i,u}=\downarrow$, and if $c_{i,u}$ changes  it is to be incremented by $1$. If $0<d_u<\floor{\frac k2}$, for any $i\ge d_u$, if $b_{i,u}=\downarrow$, the only state change can be that $b_{i,u}$ becomes $\uparrow$. If $b_{i,u}=\uparrow$, the only state change can be that $b_{i,u}$ becomes $\downarrow$ and $c_{i,u}$ gets incremented by $1$. Thus, we can consider the only possibility for a state change of a given pair $(b_{i,u}, c_{i,u})$.
%
%
%
%
%
%
We  exhaustively look at all the  possible transitions in $N(u)\cup\{u\}$ to check that we still have $\bc{u}$ and $\welld{u}$.
\begin{itemize}
    \item Since we can only apply stationary rules from $C$ (Lemma~\ref{lem:legitimate:prop}),
    all the local variables  $d_u$ remain constant:  $\forall u,\  d_u(C)=d_u(C')$. Hence, for all nodes $u$, predicate $\welld{u}$ is true in $C'$, and $S(C)=S(C')$.
    \item Let $u$ be a node. Let $X\subseteq N(u)\cup\{u\}$ be the set of nodes activated between  $C$ and  $C'$. We need to prove that in  $C'$, predicate $\bc u$ is true.
    \begin{itemize}
        \item If $d_u\ge\ksurdeux$, predicate $\bc{u}$ in $C'$ is always true, as $d_u(C')=d_u(C)$.
        \item If $d_u=0$, let $i\ge0$. We need to prove that predicate $\bcd{u}{i}$ is true in $C'$. We know, by the fact that $d_u=0$ and $leader\_down$, that $b_{i,u}=\downarrow$. Let $v\in N(v)$. Since predicate $\welld u$ is true, we have $d_v=1$. By $\bcd u i$, we have $(b_{i,v},c_{i,v})\in \{(\uparrow,c_{i,u}),(\uparrow,c_{i,u}-1), (\downarrow,c_{i,u})\} $ in~$C$. 
        
        If $c_{i,u}(C')=c_{i,u}(C)+1$, rule \textbf{Incr Leader} has been activated. It means that $\forall v\in N(u)$, $b_{i,v}(C)=\uparrow$ and $c_{i,v}(C)=c_{i,v}$. Rule \textbf{Sync 1 Down} could not have been activated on $v$ for $i$ as they had the same $c_{i,u}$. Hence    $b_{i,v}(C')=b_{i,v}(C)=\uparrow$, and $c_{i,v}(C')=c_{i,v}(C)=c_{i,u}(C)=c_{i,u}(C')-1$. Predicate $\bcd u i$ is true in~$C'$.
        
         If $c_{i,u}(C')=c_{i,u}(C)$, let us look on what can have happened to the neighbor $v$ depending on their values of pair ($b_{i,v},c_{i,v}$) in $C$.
         \begin{itemize}
             \item $ (b_{i,v},c_{i,v}) = (\uparrow,c_{i,u})$: the states will not change after one transition, as $u$'s value of local variable $c_{i,u}$ should have been one more.
             \item $(b_{i,v},c_{i,v}) = (\uparrow,c_{i,u}-1)$: either $v$ is not activated, and the state has not changed, either $v$ was activated, and Rule \textbf{Sync 1 down} was applied. In that case, the new state in $C'$ is $(\downarrow,c_{i,u})$.
             \item $(b_{i,v},c_{i,v}) =  (\downarrow,c_{i,u})$: if $v$'s state has changed, it means that Rule \textbf{Sync 1+ up} has been applied. In that case, the new state of $v$ in $C'$ is $(\uparrow,c_{i,u})$.
         \end{itemize}
         In every case, the possibilities in $C'$ for $v$ are compatible with the state of $u$.
        \item If $0<d_u\le k/2$, let $i\ge d_u$. We have four possibilities, depending on if $b_{i,u}$ is $\uparrow$ or $\downarrow$, and if $u$ changes its state or not:
        \begin{itemize}
            \item $b_{i,u}=\uparrow$, and   $u$ does not change its state.
            
            Let $v\in N(u)$ be a node such that $d_v=d_u-1$. It can only be in states in $\{(\uparrow,c_{i,u}),(\downarrow,c_{i,u}),(\downarrow,c_{i,u}+1)\}$. In the two first possibilities, if $v$ changes its state, it is still compatible with the state of $u$. In the third case, $v$ could not change its state because $u$'s state is not allowing it.
            
            Let $v\in N(u)$ be a node  such that $d_v=d_u+1$. It can only be in state $(\uparrow,c_{i,u})$. It cannot change its state, because $u$'s state is not allowing it.
            \item $b_{i,u}=\uparrow$, and $u$ changes its state to $(\downarrow,c_{i,u}+1)$. 
            
            Let $v\in N(u)$ be a node such that $d_v=d_u-1$. It can only be in state $(\downarrow,c_{i,u}+1)$, as otherwise $u$ could not have changed. In that situation, $u$'s state prevents $v$ from changing its own.
            
            Let $v\in N(u)$ be a node such that  $d_v=d_u+1$. It can only be in state $(\uparrow,c_{i,u})$. It cannot change its state  because $u$'s state is not allowing it.
            \item $b_{i,u}=\downarrow$, and $u$ does not change its state.
            
            Let $v\in N(u)$ be a node such that $d_v=d_u-1$. It can only be in states $(\downarrow,c_{i,u})$.  It cannot change its state  because $u$'s state is not allowing it.
            
            Let $v\in N(u)$ be a node such that $d_v=d_u+1$. It can only be in states in $\{(\uparrow,c_{i,u}-1),(\downarrow,c_{i,u}),(\uparrow,c_{i,u})\}$. In the two first possibilities, if $v$ changes its state, it is still compatible with the state of $u$. In the third case, $v$ could not change its state because $u$'s state is not allowing it.
            \item $b_{i,u}=\downarrow$, and $u$ changes its state.
            
            Let $v\in N(u)$ be a node such that $d_v=d_u-1$. It can only be in state $(\downarrow,c_{i,u})$.  It cannot change its state  because $u$'s state is not allowing it.
            
            Let $v\in N(u)$ be a node such that $d_v=d_u+1$. It can only be in state $(\uparrow,c_{i,u})$, as otherwise $u$ could not have changed. In that situation, $u$'s state prevents $v$ from changing its own state.
        \end{itemize}
    \end{itemize}
\end{itemize}

\end{proof}\end{articlelong}

\subsection{Reaching a Legitimate Configuration}
The goal of the following lemmas is to prove that, from any configuration $C$, we can reach a configuration $C'$ that is legitimate. The Gouda daemon's property concludes that a legitimate configuration will always eventually be reached. Indeed, let $C$ be a configuration that is infinitely often reached during an execution. Under the Gouda daemon, as a legitimate configuration $C'$ is reachable from configuration $C$, $C'$ will also be reached infinitely often.

To that end, we introduce the notion of \motnouveau{locally legitimate} node for leaders
satisfying conditions close to the legitimate ones in their ball of radius $k-1$. We prove that if a node $s$ is locally legitimate, then it will remain so forever (Lemma~\ref{lem:llremainsll}).

We explain how to make locally legitimate a node with no leader at a distance smaller than $k$ to it in Lemmas~\ref{lem:augmentLLC}~and~\ref{lem:nocollision}.
We explain how, when some leaders are too close to each other, we can reach a configuration where none of the remaining ones are at distance smaller than $k$ from another (Lemma~\ref{lem:nocollision}). 

From here, we can conclude with the proof of the following theorem:

\begin{theorem} \label{th:convergencesynchronous}
 Under the Gouda daemon, any execution eventually  reaches a legitimate configuration.
\end{theorem}

We first introduce the notion we will use in this section for nodes in $S(C)$:

\begin{definition}\label{def:locallegitimate}
Let $C$ be a configuration. A node
$s$ in $S(C)$ is  \motnouveau{locally legitimate} if
  \begin{enumerate}
\item   all the nodes  $u$ in $\mathcal B (s, \ksurdeux )$ are such that $well\_defined(u)$, $leader\_down(u)$ and\\ $branch\_coherence(u)$ hold and $d_{u} = dist(u,s) $;
\item   all the nodes  $u$ in $\mathcal B (s,k-1) \setminus \mathcal B (s, \ksurdeux )$ are such that $k-dist(u,s) \leq d_{u} \leq dist(u,s) $. 
\end{enumerate}
We denote $\mathcal{LL}(C)$ the set of those nodes in $C$.
\end{definition}

Let $s$ be a locally legitimate node. The first property means that in its neighbourhood at distance at most $\ksurdeux$, nodes behave like in a legitimate configuration. Therefore, they cannot detect errors. The second property implies that all nodes in  $\mathcal B (s,k-1)$ have coherent $d$-values according to $s$ and to potential leaders  at distance at least $k$ from $s$. A direct observation is the following:

\begin{lemma}\label{lem:llIsolated}
Let $s\in\mathcal{LL}(C)$. We have $\mathcal{B}(s,\KAmoinsUn)\cap S(C)=\{s\}$. 
\end{lemma}
\begin{articlelong}\begin{proof}
 
 
From the definition of local legitimacy, every node in $\mathcal{B}(s,\KAmoinsUn)\setminus \quickset{s}$ has positive $d$-value.
\end{proof}\end{articlelong}
Combining Lemma~\ref{lem:llIsolated} and the first property of the legitimated node, we can deduce that once a node is legitimate, it remains legitimate during the rest of the execution.

\begin{lemma}\label{lem:llremainsll}
Let $C$, $C'$ be two  configurations such that  $C\rightarrow C'$. We have $\mathcal{LL}(C) \subset  \mathcal{LL}(C')$.


\end{lemma} 
\begin{articlelong}\begin{proof}
Suppose that node $s$ is locally legitimate in $C$. Let us prove that if $C\rightarrow C'$, then $s$ is also locally legitimate in $C'$.

\begin{itemize}
    \item 
    Since $s$ is locally legitimate in $C$, the configuration obtained by restricting $C$ to nodes $\mathcal{B}(s,\ksurdeux)$ is a legitimate configuration. By Lemma~\ref{lem:legitimate:rule}, only stationary rules can be applied those nodes in that restricted configuration. Since nodes in $\mathcal{B}(s,\ksurdeux-1)$ can only see nodes in $\mathcal{B}(s,\ksurdeux)$, the rules enabled for those nodes are the same in the restricted configuration and $C$. Thus only stationary rule can be enabled on $\mathcal{B}(s,\ksurdeux-1)$ in $C$.
    
    \item For nodes at distance $\ksurdeux$ from $s$, 
    since their $d$-value is exactly $\ksurdeux$, they cannot execute neither rules \textbf{Error Spread}, \textbf{Become Leader}, \textbf{Two Heads}, \textbf{Branch incoherence} or any stationary rule. They also have at least one neighbor at distance $\ksurdeux-1$ from $s$ with $d$-value $\ksurdeux-1$. Moreover, we have these two following points:
    \begin{enumerate}
        \item The neighbors of those nodes in $\mathcal{B}(s,\ksurdeux)$ have $d$-value equal to their distance to $s$, \motnouveau{i.e.} $\ksurdeux$ or $\ksurdeux-1$;
        \item The neighbors of those nodes in $\mathcal{B}(s,k-1) \setminus \mathcal{B}(s,\ksurdeux)$ have distance to  $s$ at least $\ksurdeux+1$ and from local legitimacy their $d$-value is at least $k-(\ksurdeux+1)$ which is $\ksurdeux -1$ when $k$ is even, and $\ksurdeux$ when $k$ is odd.
    \end{enumerate}
     From these facts: 
     \begin{itemize}
    \item Rule \textbf{Update distance} is not enabled on those nodes; 
    \item For rule \textbf{Remote Collision}, the only possibility for it to be 
    enabled would be the first case in the guard when $k$ is odd. Suppose then $k$ is odd and $u$ is a node at distance $\ksurdeux$ from $s$, 
    with two distinct neighbors $v,v'$ such that $d_v=d_{v'}=d_{u}-1=\ksurdeux-1$. As $k$ is odd, neighbors of $u$ in 
    $\mathcal{B}(s,k-1) \setminus \mathcal{B}(s,\ksurdeux)$ have $d$-value at least  
    $\ksurdeux$ (see Point~2 above). Hence, nodes $v,v'$ should be in 
    $\mathcal{B}(s,\ksurdeux)$.
    But local legitimacy guarantees predicate \motnouveau{branch\_coherence} on 
    every node in $\mathcal{B}(s,\ksurdeux-1)$ which guarantees that the clock 
    $c_{\ksurdeux-1,\cdot}$ of two such nodes could not be out of sync by $2$. Thus 
    rule \textbf{Remote Collision} is not enabled on $u$.
      \end{itemize}
      And finally, since those nodes are supposed well-defined, \textbf{Reset error} is not enabled. Those nodes are then not activable.
    \item For nodes in $\mathcal{B}(s,k-1) \setminus \mathcal{B}(s,\ksurdeux)$, the condition on their $d$-value prevent them from executing any stationary rule, on any of the rules \textbf{Remote Collision}, \textbf{Two Heads}, \textbf{Branch incoherence}, and \textbf{Become leader}. The only potentially enabled rules are \textbf{Update distance}, \textbf{Error Spread} and \textbf{Reset Error}.
\end{itemize}
Then, we focus on properties in configuration $C'$.
\begin{itemize}
    \item As it was done before to analyse the enabled rules in $\mathcal{B}(s,\ksurdeux-1)$, we consider the legitimate configuration obtained by restricting $C$ to nodes $\mathcal{B}(s,\ksurdeux)$. By Theorem~\ref{th:stability}, any transition from the restricted configuration would still be legitimate. Since nodes in $\mathcal{B}(s,\ksurdeux-1)$ can only see nodes in $\mathcal{B}(s,\ksurdeux)$, and since nodes of $\mathcal{B}(s,\ksurdeux) \setminus \mathcal{B}(s,\ksurdeux-1)$ were not activated in the transition, the nodes changed state in the transition from $C$ to $C'$ as they would have from the restricted configuration, thus the restriction of $C'$ to $\mathcal{B}(s,\ksurdeux)$ is a legitimate configuration, and $well\_defined(.)$ and $branch\_coherence(.)$ are still true on nodes of $\mathcal{B}(s,\ksurdeux)$ in $C'$.
    
    \item Consider $u \in \mathcal{B}(s,k-1) \setminus \mathcal{B}(s,\ksurdeux)$. If $u$ has not changed its $d$-value in the transition, there is nothing to prove. If the $d$-value of $u$ changed in $C \rightarrow C'$, it must have performed rules \textbf{Update distance} or \textbf{Become Leader} (\textbf{Reset Error} may only change $d$-value of nodes that have an original $d$-value of 0). 
    
    Rule \textbf{Become Leader} could not have been applied, as $u$ as a neighbor has a neighbor $v$ one step closer to $s$, which means that $d_v\le dist(v,s)<k-1$.
    
    Let us prove that rule \textbf{Update distance} could not have been applied either. Every $v \in \mathcal{B}(s,k-1) \setminus \mathcal{B}(s,\ksurdeux)$ is such that $k-dist(u,s) \leq d_v \leq dist(v,s)$ in $C$ from local legitimacy, and the same can be true for the nodes exactly at distance $\ksurdeux$. Then, as neighbours of $u$ are at distance at least $dist(u,s)-1$ at most $dist(u,s)+1$ from  $s$, the minimum $d$-value in the neighborhood of $u$ in  $C$ is
    \begin{itemize}
        \item at least $k - (dist(u,s)+1)$ and,
        \item at most $dist(u,s)-1$.
    \end{itemize}
    Then if $u$ has executed \textbf{Update distance} in the transition, its new $d$-value must be at least $k - (dist(u,s)+1) + 1 = k - dist(u,s)$, and at most $dist(u,s)-1+1 = dist(u,s)$, thus $\ksurdeux < d_u \leq dist(u,s)$ in $C'$.
\end{itemize}

\end{proof}\end{articlelong}

We  focus now on how to create locally legitimate nodes. First of all, we can make sure that the $d$-values of all the nodes are coherent with regards to their distance to $S(C)$:

\begin{lemma}\label{lem:goodDist}
For any configuration $C$, we can reach a configuration $C'$ such that $S(C)=S(C')$, and $d_u=\min(dist(u,S(C')),k-1)$ for every node $u$, and there is no node with $err$-value 1 among nodes with $d$-value greater than $\ksurdeux$.
\end{lemma}
\begin{articlelong}\begin{proof}
    Let us prove the following property by induction on $i< k-1$:
 
    From any configuration $C$, we can reach a configuration $C'$ where $S(C)=S(C')$, and, for all nodes $u\in V$ we have $dist(u,S(C'))\le i\iff d_u\le i\iff d_u=dist(u,S(C'))$.
    
    \begin{itemize}
        \item For the case $i=0$, this is true by definition of $S(C')$.
        \item Let us assume that the property is true for some $i\le k-3$.
        
        Let $u$ be a node such that $dist(u,S(C'))=i+1$. In its neighborhood, there is a node $v$ such that $dist(v,S(C'))=i$. By the induction hypothesis, $d_v=dist(v,S(C'))$. Moreover, by the induction hypothesis, there is no node $w$ in $u$'s neighborhood such that $d_w<d_v$ (otherwise, $u$ would be at distance at most $d_w+1<d_v+1=dist(u,S(C'))$).

        Hence, if $d_u\neq dist(u,S(C'))$, by activating $u$, rule \textbf{Update distance} can be executed, and $d_u$ would become $d_v+1$ as $d_v<k-1$.
        
        Let $u$ be a node such that $d_u=i+1$ and $dist(u,S(C'))>i+1$.  Since all of its neighbors are at distance greater than $i$ to $S(C')$, the induction hypothesis implies that their $d$-value is at least $i+1$. Thus if $u$ gets activated, it will execute rule \textbf{Update distance}. After that, we have  $d_u>i+1$.
        
        From the current configuration $C$, by activating all the nodes $u$ at distance $i+1$ to $S(C)$ such that $d_u\neq i+1$ and all the nodes at distance greater than $i+1$ such that $d_u=i+1$, we reach a configuration $C'$ where the property is true for $i+1$.
        
        Note that we never activate nodes such that $d_u=k-1$ and for all $v\in N(u)$, $d_v=k-1$. Hence, $S(C')=S(C)$.
    \end{itemize}
    
    The induction being verified, we can reach a configuration $C'$ where, for all nodes, $d_u< k-1\iff d_u=dist(u,S(C'))$. Hence, in this configuration, all nodes at distance at least $k-1$ have $d_u=k-1$.

\end{proof}\end{articlelong}

Let $s$ be a node at distance at least $k$ from $S(C)$. We explain how to make that node locally legitimate:

\begin{lemma}\label{lem:augmentLLC}
Let $C$ be a configuration where there exists a node $s$ such that 
$dist(s,S(C)) \geq k$. A configuration $C'$ can be reached from $C$ such that $s\in\mathcal{LL}(C')$.

\end{lemma} 
\begin{articlelong}\begin{proof}
    By applying Lemma~\ref{lem:goodDist}, we can reach a configuration $C''$ where, for each node $u\in V$, $d_u=\min(dist(u,S(C'')),k-1)$. In particular, it means that  $d_s=k-1$, and it is also the case for its neighbors. 
 
    Observe that all nodes $u$ in $\mathcal{B}(s,\ksurdeux)$ have its local value $d_u$ greater that $\ksurdeux$.
 Then, in $C''$, if one of these nodes has $err$-value  equals to $1$, it can execute rule \textbf{Reset Error} which does not change its $d$-value. We can reach a configuration with the same property as $C''$ on $d$-values, without no node having
 $err_u=1$ in $\mathcal{B}(s,\ksurdeux)$.  Now, we can hence apply rule \textbf{Become Leader} to $s$. 
    
    For any integer $i$ from $1$ to $k-2$, we do a transition where we activate nodes at distance $i$ from $s$, to reach some configuration $C'$.
    
    For the first $\ksurdeux$ steps, the nodes $u$ activated have their $d_u>\ksurdeux$ before their activation (otherwise, we would have $d_s<k-1$). For those nodes, rule \textbf{Update distance} is executed. We get, for each node $u$ at step $i$, $d_u=i=dist(u,s)$. For each $j \in \segment{d_u}{\frac{k}{2}-1}, c_{j,u}:= 0$ ; $b_{i,u} := \downarrow$. For all $i\le\ksurdeux-1$, Property 1. of Definition~\ref{def:locallegitimate} is satisfied.
    
    Let $u$ be a node at distance $\ksurdeux$ from $s$. Let $v$ be a neighbor of $u$ 
    such that $d_v=\ksurdeux-1$. We can notice that $u$ is at distance $d_v$ from $s$ (otherwise, $s$ would have been at distance $<k$ from a node in $S(C)\setminus\{s\}$). Hence $u$ also satisfies  Property 1  of Definition~\ref{def:locallegitimate}.
    
    Let $u\in \mathcal{B}(s,k-1)\setminus\mathcal{B}(s,\ksurdeux)$. By direct induction, we can see that after step $dist(u,s)-1$, $u$ has a neighbor $v$ closer to $s$ such that $d_v\le dist(v,s)=dist(u,s)-1$. After applying rule \textbf{Update distance} in step $dist(u,s)$, we get that $d_u\le d_v+1\le dist(u,s)$. Moreover, we need to prove that $d_u\ge k-dist(u,s)$. We can notice that in $C''$, as each node is such that its $d$ equals its distance to $S(C'')$, no node $v$ in $\mathcal{B}(s,k-1)\setminus\mathcal{B}(s,\ksurdeux)$ is such that $d_v< k-dist(v,s)$, otherwise $dist(s,S(C))$ would be smaller than $k$, which contradicts the premise of the lemma.
    
    Let $i\le k-2$ be the first step where a node $u$ updates its distance such that $d_u< k-dist(u,s)$. It would mean that it has a node $v$ in its neighborhood such that $d_v< k-dist(u,s)-1$ in the previous step. We have $dist(v,s)\le dist(u,s)+1$. Moreover, $v$ cannot be at a closer distance to $s$, otherwise $i$ would not be minimal. Hence, $v$ was not updated in the steps from 1 to $i-1$, meaning that $d_v=dist(v,S(C''))$. As $dist(v,s)\ge\ksurdeux$, there exists some $s'\in S(C'')$ different from $s$ such that $dist(v,s')=d_v$. We obtain that $dist(s,s')\le dist(v,s)+dist(v,s')<dist(u,s)+1+k-dist(u,s)-1<k$, which contradicts the premise of the lemma.
    
    This concludes that $s$ is locally legitimate in the last configuration after the $k-1$ steps.

\end{proof}\end{articlelong}

Now, we need to deal with leaders that are too close from each other. To do this, we introduce the function that measures the number of nodes in this situation in a configuration, and Lemma~\ref{lem:gestion:des distance bis} shows how to decrease it.

\begin{definition} \label{def:phi}
 Let $C$ be a configuration. We define $\phi(C)$ as the set of leaders in $C$ having a conflict with another one due to being at distance less than $k$ to each other, \textit{i.e.}  $\phi(C) = \quickset{u \in S(C) \mid  \exists v \in S(C)\setminus \quickset{u}, dist(u,v) < k}$.
\end{definition}


\begin{lemma}\label{lem:gestion:des distance bis}
Let $C$ be a configuration such that $\phi(C) \not= \emptyset$. There exists a node $u$ in  $\phi(C)$ and a configuration $C'$ such that we can reach  $C'$ from $C$ with $S(C') = S(C) \setminus \quickset{u}$. 
\end{lemma}

\begin{articlelong}\begin{proof}
Using Lemma~\ref{lem:goodDist}, we can reach a configuration $C^*$ from $C$ where for each node $u\in V$, $d_u=\min(dist(u,S(C^*)),k-1)$. From now, we suppose that each node $u\in V$ is such that $d_u=\min(dist(u,S(C)),k-1)$ in $C$.

Let $u$ be a node in $\phi(C)$. The definition of $\phi(C)$ states that  a node  $v$ in $S(C)\setminus \quickset{u}$
is at distance at most $k$ from $u$. Thus, $v$ is also in $\phi(C)$, and $\phi(C)$ have at least two elements.

Let $u,v\in\phi(C)$  such that $dist(u,v)$ is minimum among the pairs of distinct nodes of $\phi(C)$, and let us denote $\delta = dist(u,v)$. From definition of function $\phi$,   $ dist(u,v) = \delta < k$, and we have $\ceil{\frac{\delta}{2}} \leq \ksurdeux$ with equality when $\delta = k-1$.

Since $v$ is the closest leader to $u$ in $\phi(C)$,  every node  in $\mathcal{B}(u,\ceil{\frac{\delta}{2}})$ has its distance
to $u$ as $d$-value. Symmetrically, we can apply the same argument for nodes in $\mathcal{B}(v,\ceil{\frac{\delta}{2}})$ relatively
to~$v$.

If a node $w$ in $\mathcal{B}(u,\ceil{\frac{\delta}{2}})$ is such that $err_w=1$, then 
all its neighbors with a smaller $d$-value can execute rule \textbf{Error Spread}. By following a shortest path from $w$ to $u$, we can reach a configuration where 
$err_u=1$.
When we reach a configuration where $err_u=1$, $u$ can execute rule \textbf{Reset Error}, and after its execution, we can reach 
a configuration $C'$  where  $d_u = 1$, thus $S(C') = S(C)\setminus\quickset{u}$.


Symmetrically, using the same argument as previously,  if a node $w$ in  $\mathcal{B}(v,\ceil{\frac{\delta}{2}})$ is such that $err_w=1$,  we can reach a configuration $C'$ such that $S(C') = S(C)\setminus\quickset{v}$.

Let us then suppose that there is no node in $\mathcal{B}(u,\ceil{\frac{\delta}{2}}) \cup \mathcal{B}(v,\ceil{\frac{\delta}{2}})$ such that it $err$-value is equal to $1$. 

Consider a shortest path $P$ from $u$ to $v$. We consider two cases according to the parity of its length.
\begin{itemize}
    \item When $\delta= 2i$ with $i\in \Nat$, $P=(u=u_0, u_1, .., u_{i} = v_{i}, .., v_1, v_0=v)$.
    Nodes in $\mathcal{B}(u,i) \setminus \mathcal{B}(u,i-1)$ have $d$-value $i$ and have $err$-value $0$ by hypothesis. Thus, executing stationary rules, we can make the clock with index $i$ go to $0$ for every node of $\mathcal{B}(u,i-1)$.
    Symmetrically, executing stationary rules, we can make the clock with index $i-1$ go to $2$ for every node of $\mathcal{B}(v,i-1)$.
    
    \item When $\delta=2i+1$ with $i\in \Nat$, $P=(u=u_0, u_1, .., u_{i}, v_{i}, .., v_1, v_0=v)$.
    Note that nodes in $\mathcal{B}(u,i+1) \setminus \mathcal{B}(u,i)$ have either their $d$-value equal to $i+1$ or $i$, since it would otherwise imply that another node of $S(C)$ is closer to $u$ than $v$, they also have $err$-value $0$ by hypothesis. Then, executing stationary rules, we can make the clock with index in $i$ go to $0$ for every node of $\mathcal{B}(u,i)$.
    Symmetrically, executing stationary rules, we can make the clock with index in $i$ go to $2$ for every node of $\mathcal{B}(v,i)$.
\end{itemize}

In both cases, nodes $u_i$ can execute rule \textbf{Remote Collision} and then we can make the error propagate toward $u$ executing rule \textbf{Error Spread} and reach a configuration $C'$ where $S(C') = S(C)\setminus\quickset{u}$.
\end{proof}\end{articlelong}

Thanks to this result, we prove that we can reach a configuration $C$ such that the set of conflicting nodes is empty:

\begin{lemma}\label{lem:nocollision}
From any configuration~$C$, we can reach a configuration $C'$ such that $\phi(C')=0$.
\end{lemma}

\begin{articlelong}\begin{proof}
Lemma~\ref{lem:gestion:des distance bis} states that  we can reach a configuration $C'$ where $|\phi(C')| <|\phi(C)|$ from any configuration $C$. This lemma can be proved by applying Lemma~\ref{lem:gestion:des distance bis}  at most $|\phi(C)|$  times.
\end{proof}\end{articlelong}

Now we focus on how to make leaders locally legitimate if they do not have any other leaders at distance smaller than $k$ from them.

\begin{lemma} \label{lem:new:S}
Let $C$ and $s$ be a configuration and a node  such that $\mathcal{B}(s,\KAmoinsUn)\cap S(C)=\{s\}$. We can reach a configuration $C'$ such that $s\in \mathcal{LL}(C')$.
\end{lemma}

\begin{articlelong}
\begin{proof}
First, let us apply Lemma \ref{lem:goodDist}, to ensure that all nodes have their distance to $S(C)$ up to date.

We can assume that $s\not\in \mathcal{LL}(C)$, as otherwise we take $C'=C$. The goal in the following proof is, in each considered case, to reach a configuration $C''$ such that $s\notin S(C'')$ (and $S(C)=S(C'')\cup\{s\}$). Then, by applying Lemma \ref{lem:augmentLLC} on $C''$, we can reach a configuration in which node  $s$ is locally legitimate. This actually means that we need to reach a configuration where $d_s>0$, while no other node $u$ changes its $d_u$. 

As $s\not\in \mathcal{LL}(C)$, by definition of $\mathcal{LL}$, we have three possible scenarios: 
\begin{itemize}
    \item There exists  a node $u$ in $\mathcal{B}(s,\ksurdeux)$ such that we do not have $well\_defined(u)$. As all the distances to $S(C)$ are correct, it means that
    $err_u=1$. Let us choose such a node $u$ that minimises its distance to $s$. 
    
    If $u=s$, we can apply rule \textbf{Reset Error}, which removes directly $s$ from $S(C)$.
    
    Otherwise, there is a path $(u=u_0,u_1,\ldots u_{d_u}=s)$ from $u$ to $s$. By activating $u_i$ for $i$ from $1$ to $d_u$, rule \textbf{Error Spread} will be applied each time, putting all those nodes in an error state. After that, we activate again $s$, which will remove it from $S(C)$. 
    
    \item There exists a node $u$ in $\mathcal{B}(s,\ksurdeux)$ such that we do not have $branch\_coherence(u)$. We activate that node, which will make an error appear with rule \textbf{Branch incoherence}, and we go back to the previous case.

    \item There exists a node $u$ in $\mathcal B (s,k-1) \setminus \mathcal B (s, \ksurdeux )$ such that $ d_u<k- dist(u,s) $ or $d_{u} >dist(u,s) $. 
    We  prove that it cannot actually happens, as we have $d_u=dist(u,S(C))$ and $\mathcal{B}(s,k)\cap S(C)=\{s\}$.
    We cannot have $d_{u} > dist(u,s)$, as $d_u=dist(u,S(C))\le dist(u,s)\le k-1$. Suppose that we have $k-dist(u,s) > d_{u}$. It implies, as $dist(u,s)>\ksurdeux$, that $d_u \leq \ksurdeux<dist(u,s)$, which means that there exists some $s'\neq s$ in $S(C)$ such that $dist(u,s')=d_u$. We have 
    $$dist(s,s')\le dist(u,s)+dist(u,s') < dist(u,s)+k-dist(u,s) < k.$$
    Thus $dist(s,s') \leq k-1$, which is a direct contradiction with $\mathcal{B}(s,\KAmoinsUn)\cap S(C)=\quickset{s}$.

\end{itemize}

    Hence, in the previous scenarios, we managed to reach a configuration $C'$ such that $S(C'')=S(C)\setminus\{s\}$. From this configuration, by applying Lemma~\ref{lem:augmentLLC}, we can reach a configuration $C'$ such that $s\in \mathcal{LL}(C')$.

    We proved that we can always reach a configuration where $s$ joins $\mathcal{LL}(C)$.
\end{proof}
\end{articlelong}

Now, we can prove that the number of legitimate nodes increases during the execution up until we converge to a legitimate configuration:

\begin{articlelong}
\begin{lemma}\label{lem:increasing}
Let $C$ be a  configuration. From $C$, we can  reach a configuration $C'$ such that   
\begin{enumerate}
\item either $\mathcal{LL}(C) \subsetneq  \mathcal{LL}(C')$;
\item or $C'$ is legitimate.
\end{enumerate}
\end{lemma}
\end{articlelong}

\begin{articlecourt}
\begin{lemma}\label{lem:increasing}
Let $C$ be a  configuration. From $C$, we can  reach a configuration $C'$ such that either $\mathcal{LL}(C) \subsetneq  \mathcal{LL}(C')$ or $C'$ is legitimate.
\end{lemma}
\end{articlecourt}

\begin{articlelong}\begin{proof}

Using Lemma~\ref{lem:nocollision} we can reach a configuration $C'''$ such that $\phi(C''')=0$. Then, using Lemma~\ref{lem:goodDist} on $C'''$, we can reach $C''$ such that $S(C'')=S(C''')$ (hence $\phi(C'')= \phi(C''') = \emptyset$), and every node satisfies  $d_u=\min(dist(u,S(C'')),k-1)$. If $\mathcal{LL}(C'')$ is a $(k,k-1)$-ruling set,  we can reach  a legitimate configuration $C'$ using Lemma~\ref{lem:goodDist}. Let us then suppose it's not the case. 

\begin{itemize}
    \item If $S(C'')\setminus \mathcal{LL}(C'') \not= \emptyset$, let us consider $u \in S(C'')\setminus \mathcal{LL}(C'')$. Since $\phi(C'')= \emptyset$, we know that $\mathcal{B}(u,k-1) \cap S(C'') = \quickset{u}$,  thus using Lemma~\ref{lem:new:S} we can reach a configuration $C'$ such that $u \in \mathcal{LL}(C')$.
    \item Else, as the nodes of $S(C'')=\mathcal{LL}(C'')$ does not form a $(k,k-1)$-ruling set, there exists a node $u \not\in S(C'')$ such that $dist(u,S(C''))\geq k$. We can then apply Lemma~\ref{lem:augmentLLC} to reach a configuration $C'$ where $u \in \mathcal{LL}(C')$. 
\end{itemize}
In both cases, since $u \not\in \mathcal{LL}(C'')$, we know that $u \not\in \mathcal{LL}(C)$ as local legitimacy cannot be lost from Lemma~\ref{lem:llremainsll}. Thus $\mathcal{LL}(C) \subsetneq \mathcal{LL}(C')$.
\end{proof}\end{articlelong}

This last lemma allows us to conclude with the proof of Theorem~\ref{th:convergencesynchronous}.

\begin{articlelong}\begin{proof}{ of Theorem \ref{th:convergencesynchronous}.}
Let $C$ be a configuration that is reached infinitely often under the Gouda daemon. We  prove that $C$ is legitimate.

Indeed, by applying Lemma \ref{lem:increasing}, either $C$ is legitimate, or we can reach a configuration $C'$ such that $\mathcal{LL}(C)\subsetneq \mathcal{LL}(C')$. In the second case, by the Gouda daemon, $C'$ is reached infinitely often. By Lemma \ref{lem:llremainsll}, $\mathcal{LL}(C')$ can only increase from $C'$. Hence, we will no longer be able to reach $C$, which means that $C$ is not reached infinitely often.

Hence, $C$ is legitimate.
\end{proof}\end{articlelong}

\section[From Ruling Sets to Distance-K Colorings]{From Ruling Sets to Distance-$K$ Colorings}\label{sec:distCol} 

In this section, we focus on the \motnouveau{distance-$K$ coloring} 
problem. A distance-$K$ coloring is a coloring such that any pair of nodes cannot share a color unless they are at distance greater than $K$. If the nodes having the same color form a $(K+1,K)$-ruling set, then those nodes respect the coloring constraint.

Let choose $k>K$ for our $(k,k-1)$-ruling sets. We partition the set of nodes into  two-by-two disjoint
sets $S^{(i)}$ such that each set corresponds to nodes of the 
same color. 
We build these sets one after another. Each of these sets is a distance-$k$ independent set of the graph, which is maximal among the nodes of $V\setminus\bigcup_{j<i}S^{(j)}(C)$. These sets will be built by composing an adaptation of our $(k,k-1)$-ruling set algorithm. 
Since the maximum degree of the graph is $\Delta$,
any ball of radius  $k-1$ contains at most $\Delta^{k-1}+1$ nodes. 
Hence we can partition the nodes into $\Delta^{k}$ ruling sets (we use this majoration in order to simplify the reading of the following proofs). 

For this reason,  the distance $K$-coloring algorithm is composed of  $\Delta^{k}$ parallel algorithms, each one of them computing an adapted $(k,k-1)$-ruling set. For Algorithm~$i$ and configuration $C$, we note $S^{(i)}(C)$ (or $S^{(i)}$ if there is no ambiguity) the corresponding set $S(C)$.
Each time a node $u$ is active, it applies a rule (if it can) for each ruling set algorithm. 

It is necessary to ensure that a node belongs to exactly one ruling 
set. To perform this, we number the ruling set algorithms: we denote by
$d^{(j)}_{u}$ the local variable $d_{u}$ of $u$ of the $j$-th 
algorithm.
By convention, we  assume that $u$ belongs to the $j$-th ruling 
set (or it has color $j$) if 
$j=min\{1\leq p \leq 
\Delta^{k} \mid d^{(p)}_{u} =0 \}$. To form a 
partition with the sets, we need to reach a configuration where for each node $u$, 
$|\{i\le \Delta^k \mid  d^{(i)}_{u} =0\}|=1$.
To achieve this, we modify rule \textbf{Become Leader} and add a rule to detect if a node is a leader in different layers (for Algorithm~$j$).



{\small
\begin{tabbing}
aa\=aaa\=aaa\=aaaaaaaa\=\kill

\textbf{Become Leader$^{(j)}$ ::} (priority 1) \\
\> \textbf{if} $err^{(j)}_u=0 \wedge (d^{(j)}_u=\KAmoinsUn) \wedge \forall v \in N(u), d^{(j)}_v =\KAmoinsUn \wedge \forall p<j: d^{(p)}_{u} >0 $\\

\> \textbf{then} $d^{(j)}_u := 0$ \\ 
\hspace{1.2cm} $\forall i \in \segment{1}{\floor{\frac{k}{2}}-1}, c^{(j)}_{i,u}:=0,b^{(j)}_{i,u} := \downarrow$ \\
~\\
\textbf{Belong To Two ruling sets$^{(j)}$ ::} (priority 0) \\
\> \textbf{if} $d^{(j)}_u=0\wedge\exists p<j: d^{(p)}_u=0$\\

\> \textbf{then} $d^{(j)}_u:= 1$\\
~\\
%

\end{tabbing}}

\vspace{-0.9cm}

We also modify the predicate $well\_defined$ (for Algorithm~$j$) as follows, which impacts the definition of legitimate configuration. In particular, now, a node $u$ such that $d^{(j)}_u=k-1$ does not need to have a neighbor closer to a leader if $d^{(i)}_u=0$ for some $i<j$.

\setlength\abovecaptionskip{0pt}
{\small
\begin{tabbing}
aa\=aaa\=aaa\=aaaaaaaa\=\kill
well\_defined$^{(j)}(u) \equiv err^{(j)}_u = 0 \wedge 
\forall v \in N(u), |d^{(j)}_u -d^{(j)}_v| \leq 1 \wedge$\\
$((\forall p\le j, d^{(p)}_u>0)\vee d^{(j)}_u<k-1 \Rightarrow (\exists v \in N(u), d^{(j)}_v = d^{(j)}_u-1))\wedge (d^{(j)}_u=0\Rightarrow\forall p<j, d^{(p)}_u>0)$\\

\end{tabbing}}
\vspace{-0.6cm}

We give a new definition of \motnouveau{legitimate configuration}:

\begin{definition}
 Let $j\le \Delta^k$. 
A configuration $C$ is said to be \motnouveau{legitimate for Algorithm~$j$} if, for all $i\le j$:
\begin{enumerate}
\item for any $u$ we have $well\_defined^{(i)}(u)$, $leader\_down^{(i)}(u)$ and $branch\_coherence^{(i)}(u)$ hold;
\item for any $u\neq v$ in  $S^{(i)}(C)^2$, we have $dist^{(i)}(u,v) \geq k$.
\end{enumerate}
\end{definition}

From this, we get the following adaptation of Lemma \ref{lem:legitimate:prop}. The proof remains slightly the same, with the exception that in the case of $d^{(j)}_u=k-1$, only nodes that have not a variable $d^{(i)}_u=0$ for some $i<j$ are considered.
\begin{lemma} \label{lem:legitimatebis:prop}
Let $C$ be a legitimate configuration for Algorithm~$j$.
 \begin{itemize}
\item  For any node $u$, if for all $i< j$, $d^{(i)}_{u}>0$, we have $d^{(j)}_{u} = dist(u,S^{(j)}(C))$;
\item  For any node $u$, if $d^{(i)}_{u}=0$, for all $j>i$, we have $d^{(j)}_{u} = \min(dist(u,S^{(j)}(C)),k-1)$;
 \item $S^{(j)}(C)$ is a $(k, k-1)$-ruling set of $V\setminus\bigcup_{i<j}S^{(i)}(C)$.
\end{itemize}
\end{lemma} 

With these modifications, we have the following adaptation of Theorem \ref{th:stability}:
\begin{theorem} \label{th:stabilitybis}
For all $j\le\Delta^k$, the set of legitimate configurations for Algorithm~$j$ is closed.
 Moreover, from a legitimate configuration $C$ for Algorithm~$j$, all the $d^{(j)}$-value do not change.
\end{theorem}
\begin{articlelong}\begin{proof}
We prove this theorem by induction on $j$. 
The base case $j=1$ is proved by Theorem~\ref{th:stability}. Suppose that the property is true for some $j<\Delta^k$, and we have a  legitimate configuration~$C$  for the first $j+1$ algorithms. By induction, we know that the configurations we can reach from $C$ do not change the $d^{(i)}$-values for $i\le j$, and they are legitimate for Algorithm~$i$.

The rule \textbf{Belong To Two ruling set$^{(i)}$} cannot be applied for $i\le j+1$, as we have $well\_defined^{(i)}$. The only change that can happen is about rule \textbf{Become Leader$^{(i)}$}, which can only happen less often.

The difference from the proofs of the previous section is that we have nodes with a $d^{(j+1)}$-value that is $k-1$ without a $k-2$ in their neighborhood. We use the fact that this happens only if their $d^{(i)}$-value is equal to 0 for some $i\le j$. As this value cannot change, by the induction, $d^{(j)}$ will not change either.

Hence, we conclude that the set of legitimate configurations for Algorithm~$j+1$ is closed, concluding the proof.

\end{proof}\end{articlelong}

 The proof to reach a legitimate configuration for Algorithm~$\Delta^k$   works in the same way as the proof of Theorem \ref{th:convergencesynchronous}. We need to do it one algorithm after another, from 1 to $\Delta^k$. The main difference is that we only consider nodes that are not a leader in a smaller algorithm when we increase the set of locally legitimate nodes. This leads to the result:

\begin{theorem} \label{th:convergencesynchronousbis}
 Under the Gouda daemon
 , any execution eventually  reaches a legitimate configuration in Algorithm~$\Delta^k$. 
\end{theorem}

These two theorems lead to the main result of distance-$K$ coloring:

\begin{theorem}
 Let $k$ and $K$ be two integers such that $k>K$. Under the Gouda daemon, any execution  eventually reaches a configuration $C$ such that 
\begin{itemize}
 \item $S^{(i)}(C)=\{u: d^{(i)}_{u}=0\}$ forms a distance-$k$ MIS of $V\setminus\bigcup_{j<i}S^{(j)}(C)$ in $G$
 \item The sets $S^{(1)}(C), \dots S^{(\Delta^k)}(C)$ form a distance-$K$ coloring.
 \item Every configuration in any execution starting in $C$ verifies the two above properties with the same sets as $C$.
 \end{itemize}
\end{theorem}

\section{Solving Mendable Problems}\label{sec:mending}

In this section, we want to solve a generalisation of \motnouveau{Greedy Problems}: \motnouveau{$O(1)$-Mendable Problems}, introduced in \cite{balliu2021local}. Greedy problems, such as $\Delta+1$-coloring and Maximal Independent Set, have the property that if some of the nodes have chosen an output that is locally valid (no pair of neighbors sharing a color, no adjacent nodes selected in the set), then any single node can choose an output that will keep the global solution locally valid. In a distributed setting, we cannot do this process sequentially from one node to another, but we can do it in parallel: if a set of nodes that are far enough from each other choose their output at each step, the solution can be completed. The global solution is valid if we repeat this process until all nodes have chosen an output.
To that end, we first introduce some definitions.

\subsection{Definitions}

We call a \motnouveau{Locally Checkable Problem}  (\lcl) $\Pi$ a problem where each node can check locally that its output is compatible with its neighbours. Let $\mathcal{O}$ be the set of outputs. The output $\Gamma:V\to\mathcal{O}$ is good if and only if, for all $u\in V$, $\Gamma(u)$ is compatible with the multiset $\{\Gamma(v)\mid v\in N(u)\}$. For example, in the case of Maximal Independent Set, with $\mathcal{O}=\{0,1\}$, $1$ is compatible with $\{0^k\mid k\le\Delta\}$, and $0$ is compatible with $\{11^x0^y\mid x+y<\Delta\}$. Note that we can consider radius-$r$ neighbourhood for the compatibility in the general case, which we will not do here out of simplicity. Our results can be adapted to the general version.

Let $\mathcal{O}$ be the set of outputs, and $\Gamma^*:V\to \mathcal{O}\cup\{\bot\}$. We say that $\Gamma^*$ is a partial solution if, for any $u\in V$ such that $\Gamma^*(u)\neq\bot$, we can complete the labels of the neighbors $v$ of $u$ (i.e. give an output to the nodes $v$ such that $\Gamma^*(v)=\bot$) to make $u$ compatible with it neighbors.

A problem is \motnouveau{$T$-mendable} if, from any partial solution $\Gamma^*$ and any $v\in V$ such that $\Gamma^*(v)=\bot$, there exists a partial solution $\Gamma'$ such that $\Gamma'(v)\neq\bot$, $\forall u\neq v$, and $\Gamma'(u)=\bot\Leftrightarrow \Gamma^*(u)=\bot$, and $\forall u\in V$, $dist(u,v)>T\Rightarrow \Gamma'(u)=\Gamma^*(u)$.
%
Intuitively, we can change the output of nodes at distance at most $T$ from a node $v$ when we select the output of $v$.

The \motnouveau{\local~model} is a synchronous model where each node is given a unique identifier. As there is no limit on the size of the messages for communication, after $r$ rounds, each node knows the topology of their neighborhood at distance $r$.

\begin{theorem}\label{th:mending}[Restated Theorem 6.2 from \cite{balliu2021local}]
Let $\Pi$ be a $T$-mendable \lcl~problem. $\Pi$ can be solved in $O(T\Delta^{2T})$ rounds in the \local~model if we are given a distance-$2T+1$ coloring.
\end{theorem}

One can observe that unicity of identifiers provided by the \local~model is not necessary to solve an \lcl~problem as long as nodes do not see twice the same identifier in the run.
If we know that an algorithm runs on a graph of size at most $n$ in $r(n)=o(\log n)$ rounds, then we can have it run on any graph of size at least $n$ with a distance-$r(n)$ coloring, using those colors as the new identifiers. The algorithm will not notice that the identifiers are not unique, producing a correct output. This technique has been used, for example, in~\cite{balliu2021local, brandt2017lcl}.

Hence, for a constant $T$, we can produce a distance-$r(T)$ coloring to then use the algorithm of Theorem \ref{th:mending}.


\subsection{Solving Greedy and Mendable Problems}

The goal now is to use distance-$k$ colorings to solve other problems. Let us say we want to solve $K$-mendable problem $\Pi$ for which we already have a \local\ algorithm $\mathcal{A}$ from Theorem~\ref{th:mending} (the output of node $u$ will be denoted $out_u$). 
To that end, we first build $\mathcal A'$, a self-stabilizing version of $\mathcal A$ that solves $\Pi$ assuming unique identifiers at distance $r$.
 Then we compose $\mathcal A'$ with our distance $k$-coloring algorithm (for $k$ big enough) - described in Section~\ref{sec:distCol} - and obtain then a self-stabilizing anonymous algorithm solving $\Pi$.
To simulate $r$ rounds in the \local~model, we need to compute the graph's topology at distance $r$ for each node. To compute the output of node $u$, $\mathcal A'$ will compute the exact mapping of the ball of radius $r$ centered on $u$. From it, $\mathcal{A}'$ will provide the output $\mathcal{A}$ would produce on this ball if the colors were identifiers.

In the following, we describe how each node will compute its ball.
 If we have beforehand a distance-$2r+1$ coloring, each node will have at most one node of some given color in its neighborhood at distance $r$. Hence, each node can compute a mapping of its neighborhood at distance $r$. At the beginning, each node knows its mapping at distance 0. If all the neighbors of a node $u$ know their mapping at distance $i$, $u$ can deduce its topology up to distance $i+1$. Note that we consider only cases where $r$ does not depend on the size of the graph. 

\begin{lemma}
Let $C$ be a configuration where each node $u$ has a color $c_u$ corresponding to a distance-$2r+1$ coloring and outputs $out_u=\bot$. From this configuration, under the Gouda daemon, we will reach a configuration $C'$ where each node outputs a mapping of their neighborhood at distance $r$. 
\end{lemma}

\begin{articlelong}
\begin{proof}
The algorithm  uses the two following rules:
{\small
\begin{tabbing}
aa\=aaa\=aaa\=aaaaaaaa\=\kill
\textbf{Init ::} \\
\> \textbf{if}  $out_u=\bot$\\

\> \textbf{then} $out_u := (0,G_0(c_u))$ where $G_0(x)$ is a graph of a single node colored $x$\\ 
\\
\textbf{Merge Neighbors ::} \\
\> \textbf{if}  $out_u=(i,G_u)$ with $i<r \wedge \forall v\in N(v), \exists j\ge i, out_v=(j,G_v)$\\

\> \textbf{then} $out_u := (i+1,merge_{i+1}(\{G_v:v\in N(u)\cup\{u\}\}$ 
\end{tabbing}}

The process $merge_{i+1}(\{G_v:v\in N(u)\cup\{u\}\})$ consists in merging the mappings at distance $i$ of $u$ and the one of its neighbors to produce the mapping at distance $i+1$. This can be done unambiguously as the distance-$2r+1$ coloring ensures that if several mappings have a node of color $c$, it corresponds to a single node of $V$. Note that for the case $merge_1$, it consists in knowing the color of our direct neighbors.

Under the Gouda daemon, we can make sure that, for any $i\le r$, we can reach a configuration where all nodes have computed their mapping at distance $i$.

\end{proof}
\end{articlelong}

With this lemma and Theorem \ref{th:mending}, we can conclude to the end result of this section:

\begin{theorem}
Let $\Pi$ be an \lcl~problem with mending radius $k$, that can be solved in $r=O(k\Delta^{2k})$ rounds in the \local~model. 
Let $C$ be a configuration where each node $u$ has a color $c_u$ corresponding to a distance-$2k+1$ coloring, a color $c'_u$ corresponding to a distance-$2r+1$ coloring, and outputs $out_u=\bot$. From this configuration, under the Gouda daemon, we will reach a configuration $C'$ where each node outputs a solution to $\Pi$. 
\end{theorem}


Note that in a ball of radius $2r+1$ in a graph of maximal degree $\Delta$, there are at most $\Delta^{2r+1}$ nodes. Hence, we need $\Delta^{2r+1}$ colors. For graphs where $\Delta$ is constant, we get a constant number of colors. As we also consider constant radius $r$ for the mendability, there are a finite number of possible mappings of balls at distance $r$ using those colors. Hence, in that case, our algorithms use a finite memory that does not depend on the size of the graph.

%% file: main.bbl
\begin{thebibliography}{10}

\bibitem{afek2002local}
Yehuda Afek and Shlomi Dolev.
\newblock Local stabilizer.
\newblock {\em Journal of Parallel and Distributed Computing}, 62(5):745--765,
  2002.

\bibitem{afek486local}
Yehuda Afek, Shay Kutten, and Moti Yung.
\newblock Local detection for global self stabilization.
\newblock volume 186, page 339, 1991.

\bibitem{altisen2019introduction}
Karine Altisen, St{\'e}phane Devismes, Swan Dubois, and Franck Petit.
\newblock Introduction to distributed self-stabilizing algorithms.
\newblock {\em Synthesis Lectures on Distributed Computing Theory},
  8(1):1--165, 2019.

\bibitem{awerbuch1985}
Baruch Awerbuch.
\newblock Complexity of network synchronization.
\newblock {\em J. ACM}, 32(4):804–823, 1985.

\bibitem{awerbuch1989network}
Baruch Awerbuch, Andrew~V Goldberg, Michael Luby, and Serge~A Plotkin.
\newblock Network decomposition and locality in distributed computation.
\newblock In {\em FOCS}, volume~30, pages 364--369. Citeseer, 1989.

\bibitem{balliu2021lower}
Alkida Balliu, Sebastian Brandt, Juho Hirvonen, Dennis Olivetti, Mika{\"e}l
  Rabie, and Jukka Suomela.
\newblock Lower bounds for maximal matchings and maximal independent sets.
\newblock {\em Journal of the ACM (JACM)}, 68(5):1--30, 2021.

\bibitem{balliu2022distributed}
Alkida Balliu, Sebastian Brandt, and Dennis Olivetti.
\newblock Distributed lower bounds for ruling sets.
\newblock {\em SIAM Journal on Computing}, 51(1):70--115, 2022.

\bibitem{balliu2021local}
Alkida Balliu, Juho Hirvonen, Darya Melnyk, Dennis Olivetti, Joel Rybicki, and
  Jukka Suomela.
\newblock Local mending.
\newblock In {\em International Colloquium on Structural Information and
  Communication Complexity}, pages 1--20. Springer, 2022.

\bibitem{barenboim2018locally}
Leonid Barenboim, Michael Elkin, and Uri Goldenberg.
\newblock Locally-iterative distributed ($\delta$+ 1) -coloring below
  szegedy-vishwanathan barrier, and applications to self-stabilization and to
  restricted-bandwidth models.
\newblock In {\em Proceedings of the 2018 ACM Symposium on Principles of
  Distributed Computing}, pages 437--446, 2018.

\bibitem{barenboim2016locality}
Leonid Barenboim, Michael Elkin, Seth Pettie, and Johannes Schneider.
\newblock The locality of distributed symmetry breaking.
\newblock {\em Journal of the ACM (JACM)}, 63(3):1--45, 2016.

\bibitem{bitton2021fully}
Shimon Bitton, Yuval Emek, Taisuke Izumi, and Shay Kutten.
\newblock Fully adaptive self-stabilizing transformer for lcl problems.
\newblock {\em arXiv preprint arXiv:2105.09756}, 2021.

\bibitem{brandt2017lcl}
Sebastian Brandt, Juho Hirvonen, Janne~H Korhonen, Tuomo Lempi{\"a}inen,
  Patric~RJ {\"O}sterg{\aa}rd, Christopher Purcell, Joel Rybicki, Jukka
  Suomela, and Przemys{\l}aw Uzna{\'n}ski.
\newblock Lcl problems on grids.
\newblock In {\em Proceedings of the ACM Symposium on Principles of Distributed
  Computing}, pages 101--110, 2017.

\bibitem{censor2020derandomizing}
Keren Censor-Hillel, Merav Parter, and Gregory Schwartzman.
\newblock Derandomizing local distributed algorithms under bandwidth
  restrictions.
\newblock {\em Distributed Computing}, 33(3):349--366, 2020.

\bibitem{cole1986deterministic}
Richard Cole and Uzi Vishkin.
\newblock Deterministic coin tossing with applications to optimal parallel list
  ranking.
\newblock {\em Information and Control}, 70(1):32--53, 1986.

\bibitem{cournier2001selfPIF}
Alain Cournier, AK~Datta, Franck Petit, and Vincent Villain.
\newblock Self-stabilizing pif algorithm in arbitrary rooted networks.
\newblock In {\em Proceedings 21st International Conference on Distributed
  Computing Systems}, pages 91--98. IEEE, 2001.

\bibitem{CDV06}
Alain Cournier, St{\'e}phane Devismes, and Vincent Villain.
\newblock Snap-stabilizing pif and useless computations.
\newblock In {\em 12th International Conference on Parallel and Distributed
  Systems-(ICPADS'06)}, volume~1, pages 8--pp. IEEE, 2006.

\bibitem{dolev2000self}
Shlomi Dolev.
\newblock {\em Self-stabilization}.
\newblock MIT press, 2000.

\bibitem{dubois2011taxonomy}
Swan Dubois and S{\'e}bastien Tixeuil.
\newblock A taxonomy of daemons in self-stabilization.
\newblock {\em arXiv preprint arXiv:1110.0334}, 2011.

\bibitem{ghaffari2016improved}
Mohsen Ghaffari.
\newblock An improved distributed algorithm for maximal independent set.
\newblock In {\em Proceedings of the twenty-seventh annual ACM-SIAM symposium
  on Discrete algorithms}, pages 270--277. SIAM, 2016.

\bibitem{ghaffari2021improved}
Mohsen Ghaffari, Christoph Grunau, and V{\'a}clav Rozho{\v{n}}.
\newblock Improved deterministic network decomposition.
\newblock In {\em Proceedings of the 2021 ACM-SIAM Symposium on Discrete
  Algorithms (SODA)}, pages 2904--2923. SIAM, 2021.

\bibitem{ghaffari2017complexity}
Mohsen Ghaffari, Fabian Kuhn, and Yannic Maus.
\newblock On the complexity of local distributed graph problems.
\newblock In {\em Proceedings of the 49th Annual ACM SIGACT Symposium on Theory
  of Computing}, pages 784--797, 2017.

\bibitem{ghosh1996fault}
Sukumar Ghosh, Arobinda Gupta, Ted Herman, and Sriram~V Pemmaraju.
\newblock Fault-containing self-stabilizing algorithms.
\newblock In {\em Proceedings of the fifteenth annual ACM symposium on
  Principles of distributed computing}, pages 45--54, 1996.

\bibitem{goddard2003self}
Wayne Goddard, Stephen~T Hedetniemi, David~Pokrass Jacobs, and Pradip~K
  Srimani.
\newblock Self-stabilizing protocols for maximal matching and maximal
  independent sets for ad hoc networks.
\newblock In {\em Proceedings International Parallel and Distributed Processing
  Symposium}, pages 14--pp. IEEE, 2003.

\bibitem{gouda2001theory}
Mohamed~G Gouda.
\newblock The theory of weak stabilization.
\newblock In {\em International Workshop on Self-Stabilizing Systems}, pages
  114--123. Springer, 2001.

\bibitem{guellati2010survey}
Nabil Guellati and Hamamache Kheddouci.
\newblock A survey on self-stabilizing algorithms for independence, domination,
  coloring, and matching in graphs.
\newblock {\em Journal of Parallel and Distributed Computing}, 70(4):406--415,
  2010.

\bibitem{hedetniemi2021self}
Stephen~T Hedetniemi.
\newblock Self-stabilizing domination algorithms.
\newblock {\em Structures of Domination in Graphs}, pages 485--520, 2021.

\bibitem{henzinger2019deterministic}
Monika Henzinger, Sebastian Krinninger, and Danupon Nanongkai.
\newblock A deterministic almost-tight distributed algorithm for approximating
  single-source shortest paths.
\newblock {\em SIAM Journal on Computing}, 50(3):STOC16--98, 2019.

\bibitem{ikeda2002space}
Michiyo Ikeda, Sayaka Kamei, and Hirotsugu Kakugawa.
\newblock A space-optimal self-stabilizing algorithm for the maximal
  independent set problem.
\newblock In {\em the Third International Conference on Parallel and
  Distributed Computing, Applications and Technologies (PDCAT)}, pages 70--74.
  Citeseer, 2002.

\bibitem{kutten1999fault}
Shay Kutten and David Peleg.
\newblock Fault-local distributed mending.
\newblock {\em Journal of Algorithms}, 30(1):144--165, 1999.

\bibitem{kutten2000tight}
Shay Kutten and David Peleg.
\newblock Tight fault locality.
\newblock {\em SIAM Journal on Computing}, 30(1):247--268, 2000.

\bibitem{naor1995can}
Moni Naor and Larry Stockmeyer.
\newblock What can be computed locally?
\newblock {\em SIAM Journal on Computing}, 24(6):1259--1277, 1995.

\bibitem{panconesi1995local}
Alessandro Panconesi and Aravind Srinivasan.
\newblock The local nature of $\delta$-coloring and its algorithmic
  applications.
\newblock {\em Combinatorica}, 15(2):255--280, 1995.

\bibitem{peleg2000distributed}
David Peleg.
\newblock {\em Distributed computing: a locality-sensitive approach}.
\newblock SIAM, 2000.

\bibitem{rozhovn2020polylogarithmic}
V{\'a}clav Rozho{\v{n}} and Mohsen Ghaffari.
\newblock Polylogarithmic-time deterministic network decomposition and
  distributed derandomization.
\newblock In {\em Proceedings of the 52nd Annual ACM SIGACT Symposium on Theory
  of Computing}, pages 350--363, 2020.

\bibitem{shukla1995observations}
Sandeep~K Shukla, Daniel~J Rosenkrantz, S~Sekharipuram Ravi, et~al.
\newblock Observations on self-stabilizing graph algorithms for anonymous
  networks.
\newblock In {\em Proceedings of the second workshop on self-stabilizing
  systems}, volume~7, page~15, 1995.

\bibitem{suomela2013survey}
Jukka Suomela.
\newblock Survey of local algorithms.
\newblock {\em ACM Computing Surveys (CSUR)}, 45(2):1--40, 2013.

\bibitem{turau2007linear}
Volker Turau.
\newblock Linear self-stabilizing algorithms for the independent and dominating
  set problems using an unfair distributed scheduler.
\newblock {\em Information Processing Letters}, 103(3):88--93, 2007.

\bibitem{Turau18}
Volker Turau.
\newblock Computing fault-containment times of self-stabilizing algorithms
  using lumped markov chains.
\newblock {\em Algorithms}, 11(5):58, 2018.

\bibitem{turau2019making}
Volker Turau.
\newblock Making randomized algorithms self-stabilizing.
\newblock In {\em International Colloquium on Structural Information and
  Communication Complexity}, pages 309--324. Springer, 2019.

\bibitem{turau2006randomized}
Volker Turau and Christoph Weyer.
\newblock Randomized self-stabilizing algorithms for wireless sensor networks.
\newblock In {\em Self-Organizing Systems}, pages 74--89. Springer, 2006.

\end{thebibliography}
